\documentclass[titlepage,12pt]{utarticle}

\usepackage{latexsym,amsmath,amssymb,amscd,cite}
\usepackage{epsf}
\usepackage[all]{xy}

\DeclareFontFamily{U}{rsf}{}
\DeclareFontShape{U}{rsf}{m}{n}{
  <5> <6> rsfs5 <7> <8> <9> rsfs7 <10-> rsfs10}{}
\DeclareMathAlphabet\Scr{U}{rsf}{m}{n}

\newcommand{\cO}{{\cal O}}
\newcommand{\beq}{\begin{equation}}
\newcommand{\eeq}{\end{equation}}
\newcommand{\id}{{\rm id}}
\def\lto{\longrightarrow}
\def\BC{{\mathbb C}}
\def\BZ{{\mathbb Z}}

\newcommand{\as}{{(\Ell)}}

\newcommand{\sa}{^{\as}}

\newcommand{\saa}{^{(\Ell\Ell)}}
\newcommand{\sad}{^{(\Ell\Ell')}}

\newcommand{\Ell}{\ell}
\newcommand{\sj}{^{(\Ell)}}
\newcommand{\ap}{{(a,0)}}
\newcommand{\app}{{(a,2)}}
\newcommand{\am}{{(a,1)}}
\newcommand{\amm}{{(a,3)}}

\newcommand{\bm}{{(b,1)}}
\newcommand{\minEll}{{\Ell_<}}
\newcommand{\dEll}{{\Delta}}
\newcommand{\mEll}{{\bar\Ell}}
\newcommand{\asl}[1]{{(\Ell_#1)}}
\newcommand{\asll}[2]{{(\Ell_#1 \Ell_#2)}}
 
\newcommand{\ws}{{\Sigma}}
\newcommand{\wsb}{{\partial\ws}}
\newcommand{\bnd}{{B}}
\newcommand{\RB}{{{\mathcal R}\sj_\bnd}}

\newcommand{\D}{\displaystyle}

\newcommand{\Psit}{{\tilde\Psi}}

\begin{document}        
\preprint{
  CERN--TH/2003-108\\
  {\tt hep-th/0305133}\\
}
\title{Landau-Ginzburg Realization of Open String TFT
}
\author{
  Ilka Brunner${}^a$, Manfred Herbst${}^a$, Wolfgang Lerche${}^a$,
Bernhard Scheuner${}^{a,b}$ \thanks{Work supported in part by the
Schweizerischer Nationalfonds}} 
    \oneaddress{
     ${}^a$ Theory Division\\
     CERN\\
     CH-1211 Geneva 23\\
     Switzerland\\{~}\\
     ${}^b$ Institute of Theoretical Physics\\
     University of Bern\\
     CH-3012 Bern\\
     Switzerland\\{~}\\
     \email{firstname.lastname@cern.ch}\\
    }

\Abstract{
We investigate $B$-type topological Landau-Ginzburg
theory with one variable, with $D2$-brane boundary conditions.
The allowed brane configurations are determined in terms
of the possible factorizations of the superpotential, and compute
the corresponding open string chiral rings. These are characterized
by bosonic and fermionic generators that satisfy certain relations.
Moreover we show that the disk correlators, being continuous functions of 
deformation parameters, satisfy the topological sewing constraints, thereby
proving consistency of the theory.
In addition we show that the open string LG model is, in its content,
equivalent to a certain triangulated category introduced by Kontsevich,
and thus may be viewed as a concrete physical realization of it.}

\date{May 15, 2003}

\maketitle

\section{Introduction and Summary}

The understanding of open-closed topological field theory (TFT) in
two dimensions is an important issue in string theory, for it
represents a framework to describe the vacuum structure of space-time
in the presence of $D$-branes. The works
\cite{LazaroiuTFT,MooreSegal,Moore}\ provide us with an axiomatic
definition of open-closed TFT via sewing constraints on Riemann
surfaces with boundaries, which can be given a formulation in terms
of non-commutative Frobenius algebras \cite{Natanzon}.  In a somewhat
different spirit, namely by focusing on cohomological aspects, 
2d open string TFT has been investigated for example in
\cite{HofmanMa,Hofman}\ and in \cite{mirbook,LMW}.  
A further, though closely related point of view
is based on derived categories, which is the general
mathematical framework that underlies $D$-branes \cite{Kontsevich,Douglas};
aspects of open string TFT from this perspective have been discussed,
for example, in \cite{Laza2,Laza3,AsLa,Diaconescu,AspDoug,Distler,Katz,mirbook,LazaroiuCat}.

In order to get a better understanding of the interrelation between
these more abstract viewpoints, it is desirable to investigate
concrete physical realizations of such open-closed TFT's.  First
steps were made by formulating boundary linear sigma models
\cite{GovLG,HIV,GovBFermions,HoriLinear,PM,HellermanTools,GovLSM,HellermanLSM};
these provide a framework to represent quite general $D$-brane
configurations, mathematically defined in terms of bundles and
sheaves localized on sub-manifolds, in terms of physical operators.

On the other hand, one can study boundary Landau-Ginzburg models
with superpotentials depending on continuous parameters, with the
expectation to be able to perform functionally non-trivial explicit
computations.  This is partly motivated from the experience with
the topologically twisted LG theories in the bulk, for which it is
often possible to determine the correlation functions by just using
consistency conditions \cite{DVV,CeVa1,CeVa2}.  Such Landau-Ginzburg models,
apart from being very concrete, also often allow to make direct
contact with an exactly solvable CFT description of a given theory. Moreover,
they provide a natural setting for the application of mirror symmetry
\cite{mirbook}.

So far, most approaches to a Landau-Ginzburg realization of $B$-type,
open-closed TFT's have focused on Dirichlet boundary conditions
for the fields, which correspond to $D0$-branes
\cite{GovLG,HIV,HoriLinear,MooreSegal,mirbook}.  In the present
paper we extend this line of research by working out in detail the
``minimal'' topological LG model with one variable, however with
$D2$-brane boundary conditions which turn out to provide a much
richer structure. Some general aspects of this model, as well as a 
detailed analysis for quadratic superpotentials, have been presented recently in \cite{KapustinLG}.

Specifically, we will confirm that the possible supersymmetric $D2$-brane
configurations correspond to the different possible factorizations
of the bulk superpotential. For each such brane configuration,
as well as for each pair of such configurations, we will work out the
topological open string spectrum, i.e., the boundary chiral ring.
As an important feature, this ring contains bosonic
as well as a fermionic generators, both of which satisfy certain
relations determined by the factorization data of the superpotential.

We will verify that in the unperturbed limit, the spectra match
exactly with the corresponding chiral ring elements known from the
BCFT description of B-type branes \cite{ReSc,FuSc,BH}. In the
perturbed theory, the spectrum of boundary changing operators
depends critically on the divisibility properties of certain polynomials,
and we observe an intriguing branching of
the spectrum for generic perturbations.

We will also determine a specific basis for the boundary
preserving operators, which allows to write down in an easy way
all disk correlation functions in the boundary preserving open string
sectors. Next we will demonstrate that the topological sewing constraints that
we mentioned above, are indeed satisfied by these disk correlators.
The proof applies to whole families of continuously deformed LG
theories, and involves the factorization of the superpotential and
the fermionic ring relations in a crucial manner (this goes
far beyond the usual analysis of sewing constraints which is based
on rational BCFT and which therefore applies only to a given, fixed theory).

Moreover we will make contact with a formulation of $B$-type
$D$-branes in terms of certain triangulated categories, which is
due to Kontsevich.  Extending the work of \cite{KapustinLG,orlov}, we
will in particular show that the underlying cohomology problems
are isomorphic, and thus lead to the same open string spectrum.
From this point of view, the boundary LG formulation provides a
concrete physical realization of this abstract mathematical framework.

We thus have good reasons to expect that extending our work to more
general theories will provide further insights in the relationship
between open-closed TFT and the categorial descriptions of $D$-branes,
apart from sharpening our technical tools for doing explicit computations.

{\it Acknowledgment}: We thank Albert Schwarz for discussions.

\section{$B$-type boundary conditions in LG models}

Starting with the familiar 2-dimensional $(2,2)$-supersymmetric
Landau-Ginzburg model for the bulk, one can study the effects of
introducing a world sheet boundary
\cite{Warner:1995ay,GovLG,HIV,HoriLinear,GovBFermions,KapustinLG}.
As is well known, the boundary breaks one half of the supersymmetries
and there essentially remain two types of symmetries \cite{Warner:1995ay},
which correspond to $A$- and $B$-type $D$-branes \cite{Ooguri:1996ck}.
In the present paper we will restrict ourselves to unbroken $B$-type
supersymmetry and include a boundary action such that the total
action is invariant under supersymmetry variations without imposing
any particular boundary conditions. This approach was used in
\cite{Warner:1995ay}, where it turned out that in order to achieve
this, one has to introduce a fermionic supermultiplet on the
boundary.  We will see in the next section that the boundary fermion
plays an essential role in the construction of the open string
chiral ring.\footnote{The significance of fermionic boundary ring elements has
been pointed out first in \cite{PM}.}

The main purpose of the present section is to define the physical setting
described above and to fix the notation. 

\subsection{Bulk action}

The $(2,2)$-superspace in two dimensions is spanned by two bosonic
coordinates $(x^0,x^1)$ and four fermionic coordinates
$\theta^\pm,\bar\theta^\pm$ (with $(\theta^\pm)^\dagger =
\bar\theta^\pm$). 
The supercharges and covariant derivatives are represented by
\begin{equation}
  \label{eq:Scharges}
  Q_\pm = \frac{\partial}{\partial\theta^\pm} 
        + i \bar\theta^\pm \frac{\partial}{\partial x^\pm} \: ,
\qquad
  \bar Q_\pm = - \frac{\partial}{\partial\bar\theta^\pm} 
        - i \theta^\pm \frac{\partial}{\partial x^\pm} \: ,
\end{equation}
and
\begin{equation}
  \label{eq:Sderiv}
  D_\pm = \frac{\partial}{\partial\theta^\pm} 
        - i \bar\theta^\pm \frac{\partial}{\partial x^\pm} \: ,
\qquad
  \bar D_\pm = - \frac{\partial}{\partial\bar\theta^\pm} 
        + i \theta^\pm \frac{\partial}{\partial x^\pm} \: ,
\end{equation}
where $x^\pm = x^0\!\pm\! x^1$. They satisfy the
supersymmetry algebra
\begin{equation}
  \label{eq:Salg}
  \{Q_\pm, \bar Q_\pm\} = - 2i\partial_\pm \: ,
\quad
  \{D_\pm, \bar D_\pm\} =   2i\partial_\pm \: .
\end{equation}

In the Landau-Ginzburg theory we introduce a chiral and an antichiral
superfield $\Phi$ and $\bar\Phi$, i.e., $\bar D_\pm \Phi = 0$ and
$D_\pm \bar\Phi = 0$. The expansion in component fields reads
\begin{equation}
  \Phi(y^\pm,\theta^\pm) = 
  \phi(y^\pm) + 
  \theta^+ \psi_+(y^\pm) + 
  \theta^- \psi_-(y^\pm) + 
  \theta^+\theta^- F(y^\pm) \;,\nonumber
\end{equation}
where $y^\pm = x^\pm\!\!-\!i\theta^\pm \bar\theta^\pm$. If we set
$\delta = \epsilon_+ Q_- -\epsilon_- Q_+ -
\bar\epsilon_+ \bar Q_- +\bar\epsilon_- \bar Q_+$, the variations of
the fields take the form
\begin{eqnarray}
  \label{eq:susyAux}
  \begin{array}{cc}
    \begin{array}{l@{\;=\;}l}
      \delta \phi   &   
      + \epsilon_+ \psi_- - \epsilon_- \psi_+ \;,\\[1mm]
      \delta \psi_+ & 
      + 2 i \bar\epsilon_- \partial_+ \phi + \epsilon_+ F \;,\\[1mm]
      \delta \psi_- & 
      - 2 i \bar\epsilon_+ \partial_- \phi + \epsilon_- F \;,
    \end{array} 
  \hspace{2cm} &
    \begin{array}{l@{\;=\;}l}
      \delta \bar\phi   & 
      - \bar\epsilon_+ \bar\psi_- + \bar\epsilon_- \bar\psi_+ \;, \\[1mm]
      \delta \bar\psi_+ & 
      - 2 i \epsilon_- \partial_+ \bar\phi + \bar\epsilon_+ \bar F \;, \\[1mm]
      \delta \bar\psi_- & 
      + 2 i \epsilon_+ \partial_- \bar\phi + \bar\epsilon_- \bar F \;.
    \end{array}
  \end{array}
\end{eqnarray}
In terms of the chiral and antichiral superfields one can build two
supersymmetric contributions for the action. The $D$-term is an
integral of a function $K(\Phi,\bar\Phi)$ over the full superspace.
Since we are interested only in topological quantities which do not
depend on the $D$-term,  we choose $K(\Phi,\bar\Phi)=\bar\Phi \Phi$
for simplicity. The second part is the $F$-term,
\begin{equation}
  \int_{\ws} d^2x d^2\theta\, W(\Phi)\bigr|_{\bar\theta^\pm=0} +
  \textrm{c.c.} \;.
\end{equation}
It contains the world sheet superpotential, which fully determines the
topological sector of the bulk theory. Up to total derivatives, the
bulk action can be written as 
\begin{equation}
\label{eq:Sbulk}
\begin{array}{ccc}
  S_\ws & = & 
    \D{ \int_{\ws} d^2x \left\{
     - \partial^\mu \bar\phi \partial_\mu \phi + 
     \frac{i}{2} \bar\psi_- ( \stackrel{\leftrightarrow}{\partial_0} +
     \stackrel{\leftrightarrow}{\partial_1} ) \psi_- +
     \frac{i}{2} \bar{\psi}_+ ( \stackrel{\leftrightarrow}{\partial_0} -
     \stackrel{\leftrightarrow}{\partial_1} ) \psi_+ \right. } \\
     & &  \left. \D{
     - \frac{1}{4} |W'|^2 - 
     \frac{1}{2} W'' \psi_+ \psi_- - 
     \frac{1}{2} \bar W'' \bar\psi_- \bar\psi_+
      } \right\} \: ,
\end{array}
\end{equation}
where the algebraic equation of motion $F= -1/2\,\bar W'(\bar\phi)$
was used.

\subsection{Introduction of boundary degrees of freedom}

If we wish to formulate our theory on a world sheet with boundary, one
recognizes first that the translation symmetry normal to the boundary
is broken and, therefore, also one-half of the supersymmetries are broken
\cite{Warner:1995ay,Ooguri:1996ck}. We choose the world sheet $\Sigma$ to be
given by the strip with coordinates
$(x^0,x^1) \in (\mathbb{R},[0,\pi])$. We are interested in $B$-type
supersymmetry with preserved supercharge $Q = \bar Q_+ + (-1)^S \bar Q_-$.%
\footnote{The other possibility would be $A$-type supersymmetry, with
          $Q = \bar Q_+ + (-1)^S Q_-$.
}
In terms of the parameters $\epsilon_\pm$ one can describe $B$-type
supersymmetry by setting
$\epsilon = \epsilon_+ = (-1)^{S+1}\epsilon_-$. 
For convenience we set $S=0$ and put the fermions together into
the combinations $\eta = \psi_- \!\!+\! \psi_+$ and 
$\theta = \psi_- \!\!-\! \psi_+$. Therefore, the $B$-type supersymmetry
transformations ($\delta = \epsilon \bar Q - \bar\epsilon Q$) read
\begin{equation}
  \label{eq:Bsusy}
  \begin{array}{cc}
    \begin{array}{l@{\;=\;}l}
      \delta \phi   & \epsilon \eta \;,\\[1mm]
      \delta \eta   & - 2 i \bar\epsilon \partial_0 \phi\;,\\[1mm]
      \delta \theta & 2 i \bar\epsilon \partial_1 \phi
                      + \epsilon \bar W'(\bar\phi) \;,
    \end{array}
  \hspace{2cm} &
    \begin{array}{l@{\;=\;}l}
      \delta \bar\phi   & -\bar\epsilon \bar\eta \;,\\[1mm]
      \delta \bar\eta   & 2 i \epsilon \partial_0 \bar\phi \;,\\[1mm]
      \delta \bar\theta & - 2 i \epsilon \partial_1 \bar\phi
                          + \bar\epsilon W'(\phi) \;.
    \end{array}
  \end{array}
\end{equation}
The boundary superspace is spanned by the
coordinates $\theta^0 = 1/2 (\theta^-\!\!+\!\theta^+)$ and
$\bar\theta^0 = 1/2 (\bar\theta^-\!\!+\!\bar\theta^+)$, so that the
supercharges become
\begin{equation}
  \label{eq:Schargesbound}
  \bar Q = \frac{\partial}{\partial\theta^0} 
           + i \bar\theta^0 \frac{\partial}{\partial x^0} 
	   \qquad\textrm{and}\qquad
  Q = - \frac{\partial}{\partial\bar\theta^0} 
      - i \theta^0 \frac{\partial}{\partial x^0} \;.
\end{equation}
From equations (\ref{eq:Bsusy}) we see that the fields of the chiral
multiplet $\Phi$ in the bulk rearrange into a bosonic and a fermionic 
multiplet $\Phi'$ and $\Theta'$, respectively. The bosonic superfield
$\Phi'$ containing  $\phi$ and $\eta$ turns out to be chiral,
i.e., $D\,\Phi'=0$, whereas the variation of $\Theta^\prime$ contains
the term $\partial_1 \phi$, which cannot be accomplished by
(\ref{eq:Schargesbound}). This means that $\theta$ and $F$ do not
form a chiral multiplet, but rather combine into the fermionic
superfield $\Theta'$ which satisfies
$D\,\Theta' = - 2i \partial_1\Phi'$. In components we have  
\begin{eqnarray}
  \label{eq:Sfieldsbound}
  \Phi'(y^0,\theta^0)   &=& 
     \phi(y^0)   +   \theta^0 \eta(y^0) ,
\nonumber\\
  \Theta'(y^0,\theta^0,\bar\theta^0) &=& 
     \theta(y^0) - 2 \theta^0 F(y^0) + 
     2i \; \bar\theta^0
     \bigl[\partial_1\phi(y^0) + \theta^0 \partial_1 \eta(y^0)\bigr] ,
\end{eqnarray}
where $y^0 = x^0 -i\theta^0\bar\theta^0$.

Now we turn back to the Lagrangian and construct the necessary
boundary terms in order to get a fully supersymmetric action. 
If we set $W=0$, the $B$-type supersymmetry variation of the bulk
action (\ref{eq:Sbulk}) gives rise to a surface term that can be
compensated by 
\begin{equation}
  \label{eq:Sboundpsi}
  S_{\wsb,\psi} = 
    \frac{i}{4}\int \!dx^0\,
    \left\{
      \bar\theta \eta - \bar\eta \theta
    \right\} \Bigr|_0^\pi \: .
\end{equation}
If we turn on the superpotential $W$ the following surface term:
\begin{equation}
  \label{eq:varSbulk}
  \delta (S_\ws + S_{\wsb,\psi}) =
    \frac{i}{2} \int \!dx^0\,
    \left\{
      \epsilon \bar\eta \bar W' + \bar\epsilon \eta W'
    \right\} \Bigr|_0^\pi \: 
\end{equation}
remains from the variation. It cannot be compensated by any boundary
action containing bulk fields, because the
combination $\epsilon \bar\eta$ occurs, whereas the transformations
(\ref{eq:Bsusy}) generate only $\epsilon \eta$. 

In order to ensure invariance of the action we need to introduce an
additional superfield on the boundary which is capable to compensate
(\ref{eq:varSbulk}). Following \cite{Warner:1995ay} we introduce a
boundary fermionic superfield  $\Pi$, which is not chiral but rather
satisfies: $D\,\Pi = E(\Phi')$, and which has the expansion 
\begin{equation}
  \label{eq:boundferm}
  \Pi(y^0,\theta^0,\bar\theta^0) =
     \pi(y^0) + \theta^0\,  l(y^0) - 
     \bar\theta^0 \,\bigl[E(\phi) + 
     \theta^0 \eta(y^0) E'(\phi)\bigr] \;.
\end{equation}
Its component fields transform as:
\begin{equation}
  \label{eq:BfermAux}
  \begin{array}{cc}
    \begin{array}{l@{\;=\;}l}
      \delta \pi & \epsilon l - \bar\epsilon E(\phi) \;,\\[1mm]
      \delta l   & - 2 i \bar\epsilon \partial_0 \pi
                   + \bar\epsilon \eta E'(\phi) ,
    \end{array}
  \hspace{2cm} &
    \begin{array}{l@{\;=\;}l}
    \delta \bar\pi  & \bar\epsilon \bar l 
                      - \epsilon \bar E(\bar\phi) \;,\\[1mm]
    \delta \bar l   & - 2 i \epsilon \partial_0 \bar\pi
                   - \epsilon \bar\eta \bar E'(\bar\phi) \;.
    \end{array}
  \end{array}
\end{equation}
Similar to the bulk theory we can build two terms for the action, i.e.,
\begin{equation}
  \label{eq:PiAction}
  S_\wsb  =  -\frac{1}{2} \int\!\!dx^0 d^2\theta \,
                         \bar\Pi \; \Pi \Bigr|_0^\pi\,
            -\frac{i}{2} \int_\wsb\!\!dx^0 d\theta \,
                         \Pi \; J(\Phi) \,_{\bar\theta=0}
			 \Bigr|_0^\pi
                         +\textrm{c.c.} \;.
\end{equation}
Using the algebraic equation of motion
$l=-i\bar J(\bar\phi)$, the boundary action reads
\begin{equation}
  \label{eq:Spi}
  S_\wsb  =  \int\!dx^0
            \left\{
              i \bar\pi \partial_0 \pi - 
              \frac{1}{2} \bar J J - 
              \frac{1}{2} \bar E E +
              \frac{i}{2} \pi \eta J' +
              \frac{i}{2} \bar\pi \bar\eta \bar J' -
              \frac{1}{2} \bar\pi \eta E' +
              \frac{1}{2} \pi \bar\eta \bar E'
            \right\} \biggr|_0^\pi\;,
\end{equation}
and the variation of the boundary fermion $\pi$ reduces to
\begin{equation}
  \label{eq:Bferm}
\begin{array}{ccc}
  \delta\pi & = & -i\epsilon \bar J(\bar\phi) - \bar\epsilon E(\phi) \: ,
  \\[1mm]
  \delta \bar\pi & = & i\bar\epsilon J(\phi) - \epsilon \bar E(\bar\phi)
	\: .
\end{array}
\end{equation}
We observe an invariance under the exchange of $\{\pi,E\}$ and
$\{\bar\pi,-iJ\}$, which we will use later on to choose $J$ and
$E$ such that their polynomial degrees satisfy $deg(J)\leq deg(E)$.

The kinetic term in (\ref{eq:Spi}) is supersymmetric by construction,
whereas the potential term containing $J$ is not, and this is due to the
non-chirality of $\Pi$. Rather, the transformation of (\ref{eq:Spi}) 
generates 
\begin{equation}
  \label{eq:varSbound}
  \delta S_\wsb = - \frac{i}{2} \int_\wsb \!dx^0\,
    \left\{
      \epsilon \bar\eta (\bar E \bar J)' + \bar\epsilon \eta (E J)'
    \right\} \;.
\end{equation}
But expression (\ref{eq:varSbound}) is exactly what we need in order
to compensate (\ref{eq:varSbulk}). We see that the whole action is 
invariant under supersymmetry iff \cite{KapustinLG}%
\begin{equation}
  \label{eq:WEJ}
  W = E J + {\rm const.} \;.
\end{equation}
This equation will play an essential role for the construction of the
bulk and boundary chiral rings, in that it relates the deformation
parameters of the bulk superpotential $W(\phi)$ to the parameters of
the boundary potentials $J(\phi)$ and $E(\phi)$.

\section{$B$-type $D$-branes in Landau-Ginzburg models}

\subsection{$D0$-brane boundary conditions}

So far we have not made use of any boundary conditions.  In
particular, the action constructed in the previous section is invariant
under $B$-type supersymmetry without using additional conditions for the bulk
fields on the boundary. However boundary conditions arise from the functional
variation of the action by requiring local equations of motion for
the bulk fields and have in general to form
closed orbits under supersymmetry transformations. Therefore we have
to supplement appropriate additional
conditions. The only boundary conditions which are compatible with
$B$-type supersymmetry correspond to $D0$- and $D2$-branes. In this
sub-section we will briefly recall how $D0$-branes arise in the LG
formulation found in the previous chapter. This class of $D$-branes
has already been considered in \cite{HoriLinear}, 
\cite{KapustinLG}. Subsequently, in the following section,
we will then consider $D2$-branes, which are the main focus of the present paper.

\indent The $D0$-branes are characterized by Dirichlet boundary conditions:
	\begin{equation}
  	\label{eq:Dirichlet}
	\begin{array}{ccc}
  	\textrm{D0-branes:} & \phi|_{\partial \Sigma} =
	\textrm{const.} \: , &
  	\eta|_{\partial \Sigma} = 0 \: ,
	\end{array}
	\end{equation}
and in this case the boundary fermion $\pi$ decouples from the bulk
theory, which can easily be seen from the action (\ref{eq:Spi}).
The only non-trivial fields on the boundary are $\theta$ and
$\bar\theta$. The $Q$-cohomology classes can be read off from:
	\begin{equation}
	\label{eq:QvarD}
	 \begin{array}{cc}
        \begin{array}{ccc}
        Q \phi   & = &   0 \: , \\                  
        Q \eta   & = &   2 i \partial_0 \phi \: , \\ 
        Q \theta & = & - 2 i \partial_1 \phi \: ,
        \end{array} &
        \begin{array}{ccc}
        Q \bar\phi   &=& \bar\eta \: , \\ 
        Q \bar\eta   &=& 0 \: , \\ 
        Q \bar\theta &=& -W'(\phi) \: .
        \end{array}
        \end{array}
        \end{equation}	
\noindent From the variations in the bulk we obtain the usual chiral
ring $\mathcal{R}$ of the bulk theory \cite{LVW}, which is generated by
polynomials of $\phi$ modulo $W^\prime(\phi)$,
        \begin{equation}
        \label{eq:bulk ring}
        \mathcal{R} = \frac{\mathbb{C}[\phi]}{W^\prime(\phi)} \: .
        \end{equation}
\noindent On the boundary the field $\bar\theta$ represents a
$Q$-cohomology class if $\phi$ takes its value at a critical point of
$W$ (cf. \cite{HoriLinear,KapustinLG}). Therefore, the chiral
ring ${\cal R}_B$ on the boundary is
	\begin{equation}
	\label{eq:boundary ring D}
	{\cal R}_B = 
	\frac{\mathbb{C}[\bar{\theta}]}{\bar{\theta}^2-\mathrm{const.}} \: .
	\end{equation}
\noindent In particular, the ring is independent of the choice of the
polynomials $J(\phi)$ and $E(\phi)$.

\subsection{Open string chiral rings for $D2$-brane boundary conditions}

We now turn to the more interesting $D2$-branes.%
\footnote{%
  In earlier works \cite{HIV,HoriLinear,mirbook}, 
  $D2$-branes were not much considered since
  in order to preserve half of the supersymmetries of the bulk theory,
   the superpotential $W$ was taken to be constant on $B$-type $D$-branes. We
  go here beyond this restriction because we compensate the variation
  (\ref{eq:varSbulk}) by the boundary potentials $J(\phi)$ and~$E(\phi)$. 
}
The fields $(\phi,\eta)$ in the bosonic boundary multiplet $\Phi'$
satisfy generalized Neumann boundary conditions, as follows from
the variation of the action and from consistency with supersymmetry.
This means that $\partial_1 \phi$ equals a function of $\phi$ and
$\bar\phi$ on the boundary; a similar relation holds for $\partial_1
\eta$.  An important observation is the fact that the boundary
fermion $\pi$ does not decouple. Instead, the field $\bar\theta$ is
related to it via the boundary condition
$\bar{\theta}|_{\partial \Sigma} = 
-(J' \pi + i E'\bar\pi) |_{\partial \Sigma}$.

The $Q$-cohomology classes of the topological sector on the boundary
can be extracted from (\ref{eq:Bsusy}) and (\ref{eq:Bferm}). They
in particular depend on the choice of boundary potentials via
        \begin{equation}
        \label{eq:Qvar}
        \begin{array}{cc}
        \begin{array}{ccc}
        Q \pi    & = &   E(\phi) \: ,
        \end{array} &
        \begin{array}{ccc}
        Q \bar\pi    &=& -i J(\phi) \: .
        \end{array}
        \end{array}
        \end{equation}
This means that the possible boundary spectra are determined in
terms of the possible factorizations (\ref{eq:WEJ}) of the bulk
superpotential. In the following, we will use the symbol $\as$ to 
label the various possible choices for $J(\phi)$ and $E(\phi)$,
and study for any given such choice the topological open string
spectrum. We will determine both the spectrum of boundary preserving
and boundary changing operators of a generically perturbed LG model
with one variable. For the special case of the unperturbed, i.e.
superconformal $A_{k+1}$ minimal models, we will compare the spectrum
obtained from the Landau-Ginzburg formulation with the spectrum one gets
using BCFT techniques, as reviewed in Appendix A.

Recall that the chiral ring $ \mathcal{R}$ of the bulk theory
(\ref{eq:bulk ring}) is determined \cite{LVW}\ in terms of the superpotential
$W(\phi)$. Assuming that $W(\phi)$ is of degree $k+2$, the ring may be
represented by polynomials $\Phi_i$ in $\phi$ with degrees $\deg
\Phi_i = i$ equal to or less than $k$:
        \begin{equation}
        \label{eq:bulk basis}
	\left\{ \Phi_i \right\} =
	\left\{ 1,\Phi_1(\phi),...,\Phi_k(\phi) \right\} \: .  
        \end{equation}
On the other hand, eq.\ (\ref{eq:Qvar}) implies that on the boundary
the chiral ring $\RB$ is truncated earlier since it consists
of polynomials modulo $J \sa (\phi)$ and $E \sa (\phi)$. In the
generic case, when $J \sa (\phi)$ and $E \sa (\phi)$ have no common
divisor, the $Q$-cohomology is empty and all topological boundary
amplitudes vanish. 
The interesting case is when the boundary potentials have a common
factor, so that we can write
       \begin{equation}
        \label{eq:EJfact}
        \begin{array}{cc}
        \begin{array}{ccc}
        J\sa(\phi) & = & q \sa(\phi) G\sa(\phi) \: ,
        \end{array} &
        \begin{array}{ccc}
        E\sa(\phi) & = & p \sa(\phi) G\sa(\phi) \: .
        \end{array}
        \end{array}
        \end{equation}
Here $G\sa(\phi)$ is the greatest common divisor of $J\sa(\phi)$
and $E\sa(\phi)$; if it is non-trivial, the bosonic part of the
boundary ring is given by the polynomials in $\phi$ modulo truncation by
$G\sa(\phi)$.

In contrast to the bulk, the chiral ring at the boundary also contains
fermionic fields, since we can construct the following $Q$-closed field
out of the boundary fermions $\pi$ and $\bar\pi$:
       \begin{equation}
        \label{eq:omegadef}
        \omega\saa \ =\ \sqrt{i} (q\sa(\phi) \pi - i
	p\sa(\phi)\bar\pi)\ . 
        \end{equation}
Here the labels indicate that $\omega\saa$ is a boundary preserving
operator, but we will sometimes omit these labels for
notational simplicity. There is an algebraic relation that $\omega\saa$
satisfies, and it is determined by the canonical anticommutation
relations \cite{Ishida:1979bc}
\begin{equation}
  \label{eq:canonical}
  \begin{array}{r@{\;\;=\;\;}l}
    \{\pi,\bar\pi\} & 1 \;,\\
    \{\pi,\pi\} & 0 \;=\; \{\bar\pi,\bar\pi\} \;.
  \end{array}
\end{equation}
One immediately obtains:%
\footnote{%
This holds up to a possible normalization factor, which
can be determined from the topological Cardy condition, 
as we will show below.
}
        \begin{equation}
        \label{eq:omega}
        \begin{array}{ccc}
        \left[ \omega\saa \right]^2 = p\sa(\phi) q\sa(\phi) \: .  
        \end{array}
        \end{equation}
In addition, the anticommutation relations allow us to write the
boundary part $Q_\bnd\sa$ of the supercharge as
	\begin{equation}
	\label{eq:Qrep}
	Q_\bnd\sa = E\sa \bar\pi -i J\sa \pi \: .
	\end{equation}
Note that in contrast to the total supercharge,
$Q_\bnd\sa$ is not nilpotent by itself:
\begin{equation}
  [Q_\bnd\sa{}]^2 = -i E\sj J\sj = -i(W-\mathrm{const.}) \;.
\end{equation}

The chiral ring $\RB$ in the boundary sector $(\Ell)$
is thus given by the polynomial ring generated by
$\phi$ and $\omega$, modulo $G\sa(\phi)$:
        \begin{equation}
        \label{eq:boundary ring}
        \RB\ =\
        \frac{\mathbb{C}[\phi,\omega]}{\left\{
        G\sa(\phi),\omega^2-p\sa(\phi)q\sa(\phi) \right\}}\ .    
                \end{equation}
The number of elements of the open string chiral ring is
controlled by the polynomial degree $d_\Ell = \deg (G\sa)$. In
total we have $d_\Ell$ bosonic fields and $d_\Ell$ fermionic fields
in the boundary preserving sector.
 In order to fix notation, let us denote these fields by
 $\Psi_\alpha\saa$, where $\alpha$ labels bosonic and fermionic
sub-sectors in an obvious manner:  $\alpha\equiv(a,\sigma)$ and
$a=0,1,\ldots,d_\Ell-1$, $\sigma=0,1$. We can thus write a basis
of $\RB$ as
        \begin{equation}
        \label{eq:boundary basis}
        \begin{array}{c@{\;=\;}l}
        \{\Psi_{\ap}\saa\} & \left\{ 1,\Psi_1(\phi),...,
        \Psi_{d_\Ell-1}(\phi)\right\} \ , \\[0,2cm]
        \{\Psi_{\am}\saa\} & 
        \left\{ \omega\saa,\omega\saa \Psit_1(\phi),...,
        \omega\saa \Psit_{d_\Ell-1}(\phi) \right\} \ ,
        \end{array}
        \end{equation}
where $\Psi_a(\phi),\Psit_a(\phi)$ are polynomials in $\phi$ of degree
$\deg (\Psi_a) = a$, which will in general be different from the bulk
ring polynomials $\Phi_i(\phi)$ in (\ref{eq:bulk basis}).

In order to determine the spectrum and the chiral ring
$\mathcal{R}_B^{\asll{1}{2}}$ for boundary changing fields, we can proceed
in a similar way as above. First, the action of the
supercharge $Q_\bnd$ on the boundary fields in the sector
$\asll{1}{2}$ can consistently be defined as 
	\begin{equation}
	\label{eq:Qact}
	\{ Q_\bnd,\Psi_\alpha^\asll{1}{2} \} \equiv
	Q_\bnd^\asl{1} \; \Psi_\alpha^\asll{1}{2} -(-)^{|\alpha|} \;
	\Psi_\alpha^\asll{1}{2} \; Q_\bnd^\asl{2} \: .
	\end{equation}
Then we realize that the canonical commutation relations
(\ref{eq:canonical}) for $\pi$
and $\bar\pi$ are universal for all boundary conditions, i.e., they do
not depend on the polynomials $J\sa$ and $E\sa$. In fact, the
supercharge $Q_\bnd\sa$ (\ref{eq:Qrep}) contains all the information on
the boundary condition $\as$. 
This implies that we can use the universality of (\ref{eq:canonical}) to
construct the boundary changing operators in terms of polynomials of
$\phi$, $\pi$ and $\bar\pi$.

We thus make the ansatz
$\omega\sad = \rho(\phi) \pi + \sigma(\phi) \bar\pi$ for the fermionic
boundary changing operators and determine its $Q$-cohomology using
(\ref{eq:Qact}). In order to do so, it is convenient to define the
following factorizations
	\begin{equation}
        \label{eq:bound changing fact}
        \begin{array}{cc}
        \begin{array}{ccc}
        E^\asl{1} & = & \hat p^\asl{1} \cdot
	\gcd\{ J^\asl{2} , E^\asl{1} \} \: , \\[0,1cm] 
        J^\asl{2} & = & \hat q^\asl{2} \cdot
	\gcd\{ J^\asl{2} , E^\asl{1} \} \: ,
        \end{array} &
        \begin{array}{ccc}
        E^\asl{2} & = & \hat p^\asl{2} \cdot
	\gcd\{ J^\asl{1} , E^\asl{2} \} \: , \\[0.1cm]
        J^\asl{1} & = & \hat q^\asl{1} \cdot
	\gcd\{ J^\asl{1} , E^\asl{2} \} \: .
        \end{array}
        \end{array}
        \end{equation}
When computing the $Q$-cohomology we observe that
$E^\asl{1}J^\asl{1} = E^\asl{2}J^\asl{2}$, which implies that in order
to obtain nontrivial cohomology classes, the constant
in (\ref{eq:WEJ}) must be the same for the boundary sectors $\asl{1}$
and $\asl{2}$. Moreover, from (\ref{eq:bound changing fact}) we find that
$\hat p^\asl{1}\hat q^\asl{1} = \hat p^\asl{2}\hat q^\asl{2}$. It turns
out that there occur two kinds of fermionic solutions for the
$Q$-cohomology classes, i.e.,
	\begin{equation}
	\label{eq:changing ferm}
	\begin{array}{r@{\;\;=\;\;}l}
	  \omega_{qp}^\asll{1}{2} \;\Psi_{qp}&
	  \sqrt{i} ( \hat q^\asl{1} \pi - 
	  i \hat p^\asl{2}\bar\pi) \;\Psi_{qp} \;, \\[2mm]
	  \omega_{pq}^\asll{1}{2} \;\Psi_{pq}&
	  \sqrt{i} ( \hat q^\asl{2} \pi - 
	  i \hat p^\asl{1}\bar\pi) \;\Psi_{pq} \;,
	\end{array}
	\end{equation}
where $\Psi_{qp}$ and $\Psi_{pq}$ are polynomials of $\phi$ modulo
$\gcd\{ J^\asl{1} , E^\asl{2} \}$ and $\gcd\{ J^\asl{2} , E^\asl{1}
\}$, respectively. The solutions (\ref{eq:changing ferm}) are not
completely independent but rather satisfy the relations
\begin{equation}
  \label{eq:relation omega}
  \begin{array}{r@{\;\;=\;\;}l}
    \hat p^\asl{1}\;\omega_{qp}^\asll{1}{2} & 
    \hat p^\asl{2}\;\omega_{pq}^\asll{1}{2} \;,\\
    \hat q^\asl{2}\;\omega_{qp}^\asll{1}{2} & 
    \hat q^\asl{1}\;\omega_{pq}^\asll{1}{2} \;,\\
  \end{array}
\end{equation}
where it is clear that common divisors could be divided out. 
In a similar way we make the ansatz
$\beta\sad = \rho(\phi) \pi\bar\pi + \sigma(\phi) \bar\pi\pi$ for the
bosonic boundary changing operators. We define the following
factorizations appropriate for this case:
        \begin{equation}
        \label{eq:bound changing fact bos}
        \begin{array}{cc}
        \begin{array}{ccc}
        E^\asl{1} & = & e^\asl{1} \cdot \gcd\{ E^\asl{1} ,
	E^\asl{2} \} \: , \\[0.1cm]
        E^\asl{2} & = & e^\asl{2} \cdot \gcd\{ E^\asl{1} ,
	E^\asl{2} \} \: ,
        \end{array} &
        \begin{array}{ccc}
        J^\asl{2} & = & j^\asl{2} \cdot \gcd\{ J^\asl{1} ,
	J^\asl{2} \} \: , \\[0.1cm]
        J^\asl{1} & = & j^\asl{1} \cdot \gcd\{ J^\asl{1} ,
	J^\asl{2} \} \: ,
        \end{array}
        \end{array}
        \end{equation}
which imply $e^\asl{1}j^\asl{1}=e^\asl{2}j^\asl{2}$. Likewise, there
exist two kinds of solutions for the boundary changing bosons, which
can be written as 
	\begin{equation}
	\label{eq:changing boson}
	\begin{array}{r@{\;\;=\;\;}l}
	\beta_j^\asll{1}{2} \;\Psi_j & 
	(j^\asl{1} \pi\bar\pi + j^\asl{2} \bar\pi\pi) \;\Psi_j \;,\\[0.2cm] 
	\beta_e^\asll{1}{2} \;\Psi_e & 
	(e^\asl{2} \pi\bar\pi + e^\asl{1} \bar\pi\pi) \;\Psi_e \;,
	\end{array} 
	\end{equation}
$\Psi_j$ and $\Psi_e$ being polynomials modulo
$\gcd\{ J^\asl{1} ,J^\asl{2} \}$ and 
$\gcd\{ E^\asl{1} ,E^\asl{2} \}$, respectively. We have again
relations between the solutions (\ref{eq:changing boson}), namely
\begin{equation}
  \label{eq:relation beta}
  \begin{array}{r@{\;\;=\;\;}l}
    e^\asl{1}\;\beta_j^\asll{1}{2} & 
    j^\asl{2}\;\beta_e^\asll{1}{2} \;,\\[1mm]
    e^\asl{2}\;\beta_j^\asll{1}{2} & 
    j^\asl{1}\;\beta_e^\asll{1}{2} \;.
  \end{array}  
\end{equation}

Summarizing, what we have found is, in contrast to the boundary preserving sector, that the spectrum ``doubles'' into two sets of bosonic and two sets of fermionic fields (at least for sufficiently generic perturbations).
For a given sector $\asll{1}{2}$ we can represent it in the following 
manner, modulo the relations (\ref{eq:relation omega}) and
(\ref{eq:relation beta}):
        \begin{equation}
        \label{eq:changing basis}
	\begin{array}{r@{\;\;=\;\;}l}
        \Psi_{\ap}^\asll{1}{2} & \bigl\{
	\beta_j^\asll{1}{2},\beta_j^\asll{1}{2} \Psi_1(\phi),...,
        \beta_j^\asll{1}{2} 
	\Psi_{d_j\!-1}(\phi) \bigr\} \: , \\[0.2cm]
        \Psi_{\app}^\asll{1}{2} & \bigl\{
	\beta_e^\asll{1}{2},\beta_e^\asll{1}{2} \Psi_1(\phi),...,
        \beta_e^\asll{1}{2} 
	\Psi_{d_e\!-1}(\phi) \bigr\} \: , \\[0.2cm]
        \Psi_{\am}^\asll{1}{2} & \bigl\{
	\omega_{qp}^\asll{1}{2},\omega_{qp}^\asll{1}{2} \Psi_1(\phi),...,
        \omega_{qp}^\asll{1}{2} 
	\Psi_{d_{qp}\!-1}(\phi) \bigr\} \: , \\[0.2cm]
        \Psi_{\amm}^\asll{1}{2} & \bigl\{
	\omega_{pq}^\asll{1}{2},\omega_{pq}^\asll{1}{2} \Psi_1(\phi),...,
        \omega_{pq}^\asll{1}{2}
	\Psi_{d_{pq}\!-1}(\phi) \bigr\} \: ,
	\end{array}
        \end{equation}
where the $d$'s are the polynomial degrees of the respective
divisors. In (\ref{eq:changing basis}) we have extended the set of
possible values of the index $\sigma$ in the boundary changing
sectors to $\{0,1,2,3\}$, in order to account for the enlarged
spectrum. Note that the actual spectrum for a given pair of
factorizations is governed by which subsets of roots are common to
which factors, and under specific circumstances, an example for
which we will discuss momentarily, the  basis (\ref{eq:changing
basis}) may collapse to a smaller one.

For the remainder of this section, let us discuss the 
unperturbed theory, which corresponds to the twisted $N=2$
minimal model with homogeneous superpotential of singularity type
$A_{k+1}$:
\begin{equation}
W_{k+2}(\phi)\
=\ \frac{1}{k+2}\,\phi^{k+2}.
\end{equation}
This theory has an unbroken $U(1)$ R-symmetry, and in order to
maintain it on the boundary, we require $J(\phi)$ and $E(\phi)$ to be
homogeneous as well. Equation (\ref{eq:WEJ}) restricts the degrees of
$J(\phi)$ and $E(\phi)$ to certain possibilities, and by an exchange
of $\{\pi,E\}$ and $\{\bar\pi,-iJ\}$ we can always choose
$deg(J) \leq deg(E)$. All-in-all we have the following possibilities:%
\footnote{
  The choice $J(\phi) = 1$ and 
  $E(\phi) = \frac{1}{k+2}\phi^{k+2}$ was excluded,
  because in that case a constant would already be $Q$-exact and   
  all topological correlators would vanish.
}%
	\begin{equation}
	\begin{array}{ccc}
	J\sj(\phi)= \phi^{\Ell+1} \: , & E\sj(\phi)=\D{\frac{1}{k+2}}
	\phi^{k+1-\Ell} \: , & \mbox{for $\Ell \in \{0,1,...,[k/2]
	\}$} \: . 
	\end{array}
	\end{equation}
This indeed reproduces the set of $B$-type
boundary labels in the rational boundary CFT of type $A_{k+1}$, 
as reviewed in Appendix \ref{sec:CFT}.
\begin{table}[t]
\begin{center}
\begin{tabular}{|c|c||c|c|}
\hline
bosons                                         & $q\,(k+2)$        & 
fermions                                       & $q\,(k+2)$        \\
\hline
\vspace{-4mm} & & &       \\
$\Psi\saa_{(0,0)} = 1$                                & $0$               &
$\Psi\saa_{(0,1)} = \omega\;\;$                      & $k-2\Ell$   \\[2mm]
$\Psi\saa_{(1,0)} = \phi$                             & $2$               & 
$\Psi\saa_{(1,1)} = \omega\phi$                      & $k-2\Ell+2$ \\
\vdots                                         & \vdots            & 
\vdots                                         & \vdots            \\
$\Psi\saa_{(n,0)} = \phi^n$                           & $2n$              & 
$\Psi\saa_{(n,1)} = \omega\phi^n$                    & $k-2\Ell+2n$\\
\vdots                                         & \vdots            & 
\vdots                                         & \vdots            \\
$\Psi\saa_{(\Ell,0)}=\phi^{\Ell}$       & $2\Ell$     & 
$\Psi\saa_{(\Ell,1)}=\omega\phi^{\Ell}$& $k$               \\[1mm]
\hline
\end{tabular}
\caption{
  \label{tab:charges1}
  Elements of the boundary preserving chiral ring
  and their charges. They match precisely the open string states
  obtained from BCFT.}
\end{center}
\end{table}
Moreover, we can also precisely match the spectrum of boundary
fields for any given such boundary condition labeled by $(\Ell)$. For this,
recall that the charge of the bulk field $\phi$ is determined from
the bulk potential, whereas the charge of the boundary fermion
$\pi$ follows from the boundary potential in (\ref{eq:PiAction}),
i.e., $q_\phi=-q_{\bar\phi}=\frac{2}{k+2}$ and
$q_\pi=-q_{\bar\pi}=\frac{k-2\Ell}{k+2}$ (we used here the
fact that on the boundary the $U(1)$-charge is the sum of left and
right charges in the bulk). Furthermore, the $Q$-closed fermion
$\omega\saa$ takes the form
	\begin{equation}
  	\label{eq:omegaAk1}
  	\omega\saa = \sqrt{i}(\pi - \frac{i}{k+2}\,\phi^{k-2\Ell}\bar\pi),
	\end{equation}
and it has the same charge as $\pi$; it obviously satisfies the relation
(\ref{eq:omega}): $[\omega\saa]^2=\frac{1}{k+2}\phi^{k-2\Ell}$. 
Together with
$\phi$ it generates the boundary chiral ring, and from $U(1)$
conservation we get that the natural basis is very simple:
$\Psi_a(\phi)=\phi^a$, i.e.
        \begin{equation}
        \label{eq:basisAk1}
        \begin{array}{c@{\;=\;}l}
        \{\Psi_{\ap}\saa\} & \left\{ 1,\phi,...,\phi^\Ell \right\} \: ,
	\\[0,1cm] 
        \{\Psi_{\am}\saa\} & \left\{ \omega\saa,\omega\saa
	\phi,...,\omega\saa \phi^\Ell \right\} \: .
        \end{array}
        \end{equation}
\begin{table}[t]
\begin{center}
\begin{tabular}{|l@{\;=\;}l|c||l@{\;=\;}l|c|}
\hline
\multicolumn{2}{|c|}{bosons}                     & $q\,(k+2)$        & 
\multicolumn{2}{|c|}{fermions}                   & $q\,(k+2)$        \\
\hline
\multicolumn{2}{|c|}{\vspace{-4mm}}                     &      & 
\multicolumn{2}{|c|}{}                   &       \\
$\Psi^\asll{1}{2}_{(\dEll,0)}$     & $\beta^\asll{1}{2}$         & $2\dEll$    &
$\Psi^\asll{1}{2}_{(\dEll,1)}$     & $\omega^\asll{1}{2}\;\;$    & $k-2\,\mEll$        \\[2mm]
$\Psi^\asll{1}{2}_{(\dEll+1,0)}$ & $\beta^\asll{1}{2}\phi$     & $2(\dEll\!+\!1)$  & 
$\Psi^\asll{1}{2}_{(\dEll+1,1)}$ & $\omega^\asll{1}{2}\phi$    & $k-2(\mEll\!-\!1)$  \\
\multicolumn{2}{|c|}{\vdots}                     & \quad\vdots       & 
\multicolumn{2}{c|}{\vdots}                      & \quad\vdots       \\
$\Psi^\asll{1}{2}_{(\dEll+n,0)}$ & $\beta^\asll{1}{2}\phi^n$   & $2(\dEll\!+\!n)$ & 
$\Psi^\asll{1}{2}_{(\dEll+n,1)}$ & $\omega^\asll{1}{2}\phi^n$  & $k-2(\mEll\!-\!n)$ \\
\multicolumn{2}{|c|}{\vdots}                     & \quad\vdots       & 
\multicolumn{2}{c|}{\vdots}                      & \quad\vdots       \\
$\Psi^\asll{1}{2}_{(\mEll,0)}$     & 
$\beta^\asll{1}{2}\phi^{\minEll}$                        & $2\,\mEll$  & 
$\Psi^\asll{1}{2}_{(\mEll,1)}$     & 
$\omega^\asll{1}{2}\phi^{\minEll}$                       & $k - 2\dEll$
\\[1mm] 
\hline
\end{tabular}
\caption{
  \label{tab:chargesb2}Elements of the boundary changing chiral rings
  and their charges ($\dEll = \frac{1}{2}|\Ell_1\!-\!\Ell_2|$,
  $\mEll=\frac{1}{2}(\Ell_1\!+\!\Ell_2)$ and
  $\minEll = {\rm min}\{\Ell_1 ,\Ell_2\}$). 
  These match as well the results from BCFT.}
\end{center}
\end{table}
In the boundary changing sector $\asll{1}{2}$, the generators of the
algebra read
	\begin{equation}
	\label{eq:changingAk1}
	\begin{array}{c@{\;=\;}l}
	\beta^\asll{1}{2} & \left\{
	\begin{array}{l@{\quad:\quad}l}
	\phi^{\Ell_1-\Ell_2} \pi\bar\pi + \bar\pi\pi & \Ell_2 \leq
	\Ell_1 \\[0,1cm] 
	\pi\bar\pi + \phi^{\Ell_2-\Ell_1} \bar\pi\pi & {\rm otherwise}
	\end{array}  
	\right. \: , \\[0.5cm]
	\omega^\asll{1}{2} & \sqrt{i} \left(\pi - \D{\frac{i}{k+2}}
	\phi^{k-\Ell_1 -\Ell_2} \bar\pi \right) \: .
	\end{array}
	\end{equation}
From (\ref{eq:changingAk1}) we find the intriguing feature that in
this degenerate situation, the two sorts of each bosonic and
fermionic fields (\ref{eq:changing basis}) reduce to only one kind
of bosons and fermions, respectively; in other words, the basis
collapses to

        \begin{equation}
        \label{eq:changing basisAk1}
	\begin{array}{c}
        \Psi_{\ap}^\asll{1}{2} = \left\{
	\beta^\asll{1}{2},\beta^\asll{1}{2} \phi,...,
        \beta^\asll{1}{2} \phi^\minEll  \right\} \: , \\[0.2cm]
        \Psi_{\am}^\asll{1}{2} = \left\{
	\omega^\asll{1}{2},\omega^\asll{1}{2} \phi,...,
        \omega^\asll{1}{2} \phi^\minEll \right\} \: ,
	\end{array}
        \end{equation}
where $\minEll = {\rm min}\{\Ell_1 ,\Ell_2\}$.
At first sight the charge of boundary changing operators is not obvious,
because $\pi$ and $\bar\pi$ do not have a well defined charge in that 
case. However, taking advantage of charge conservation and the
operator product $\beta^\asll{1}{2} \, \beta^\asll{2}{1} =
\phi^{|\Ell_1 - \Ell_2|}$ mod $\phi^{\minEll+1}$ as well as  
$\omega^\asll{1}{2} \,\omega^\asll{2}{1} = \phi^{k-\Ell_1-\Ell_2}$ mod
$\phi^{\minEll+1}$, we conclude that
$q(\beta^{\asll{1}{2}})=\frac{|\Ell_1-\Ell_2|}{k+2}$ and  
$q(\omega^\asll{1}{2})=\frac{k-\Ell_1-\Ell_2}{k+2}$.

The R-charges for the boundary fields in the basis
(\ref{eq:basisAk1}) and (\ref{eq:changing basisAk1}) are listed in
Tables \ref{tab:charges1} and \ref{tab:chargesb2}, respectively;  
as can be inferred from Appendix \ref{sec:CFT}, they perfectly coincide with the
charges of the boundary preserving and boundary changing, 
B-type open string states of the
BCFT description of the $A_{k+1}$ minimal model.

\subsection{Disk correlators and families of deformed LG theories} 

We now return to discussing the perturbed LG theory.  For simplicity,
we will focus on the factorization preserving deformations of a
given boundary sector labelled by $(\Ell)$, while leaving
the generalization to boundary changing sectors to future work.
Moreover, we will restrict the discussion to factorizations with
$q(\phi)=1$ (so that $J\sj=G\sj$), the generalization to general
$q$ being straightforward.

We will thus consider a perturbed bulk superpotential of the form
\begin{equation}
       \label{eq:pertW}
 W_{k+2}(\phi,t)\ =\ \frac1{k+2}\,\phi^{k+2} -
  \sum_{i=0}^{k} g_{k+2-i}(t)\,\phi^{i}\  ,
\end{equation}
in connection with the following perturbation at the boundary:
\begin{equation}
       \label{eq:pertG}
 G\sj(\phi,u)\ =\ \phi^{\Ell+1}-
  \sum_{a=0}^{\Ell} h_{\Ell+1-a}(u)\,\phi^{a}\ ,
\end{equation}
where $g_{k+2-i}(t)=t_{k+2-i}+O(t^2)$, $h_{\Ell+1-a}(u)=u_{\Ell+1-a}+O(u^2)$
are certain polynomials of the flat bulk and boundary coordinates
(note that $u_1$ is an allowed deformation because the ring truncates
at the boundary in a different manner as in the bulk).
From the factorization condition (\ref{eq:WEJ}) it is clear that
the boundary deformation parameters $u_{\Ell+1-a}$ are not independent
from the bulk parameters $t_{k+2-i}$, 
rather these two sets of parameters will need to be locked together.
That is, if we write (where, as we said, we will put $q\sj(\phi,v)=1$
for simplicity):
\begin{eqnarray}
  \label{eq:tufactor}
  E\sj &=& p\sj(\phi,u,v)\,G\sj(\phi,u)\, ,
  \nonumber\\
 J\sj &=& q\sj(\phi,u,v)\,G\sj(\phi,u) \, ,
\end{eqnarray}
then the condition
$W(\phi,t)=E\sa(\phi,u,v)\,J\sa(\phi,u,v)+\mathrm{const.}\ $ determines
the bulk parameters $t_{k+2-i}$ in terms 
of the boundary parameters $u_{\Ell+1-a}$ (plus some possible extra
parameters, $v_n$). Obviously,
from the bulk point of view, when $G$ is non-trivial
the parametrization $t(u,v)$ represents
a specialization to a sub-manifold of the parameter space and
implies certain relations among the $t$'s; thus, the theory is best
parametrized by $u,v$. 

Our aim is to study the dependence of correlators as functions of
the deformation parameters, and for this it suffices to study
the dependence of the various operator product coefficients, i.e.,
the structure constants of the boundary ring.  By standard TFT
arguments, these can be computed by simple polynomial multiplication
(the fields forming the basis (\ref{eq:bulk basis}) and (\ref{eq:boundary
basis}) are $Q$-closed and therefore any correlation function is
independent from the position of the insertions).  More precisely,
in terms of the notation introduced above, the various bulk and
boundary operator products, as well as the bulk-boundary couplings
$e$, are defined as follows:%
\footnote{%
Note that the operator product $\Phi_i\,\Psi_\alpha\saa$ is not
fundamental, as it is determined, via sewing constraints, in terms
of $B$ and $e$.
}
	\begin{equation}
        \label{eq:structure}
        \begin{array}{rclcll}
	C(\Phi_i ,\Phi_j) &=& 
        {C_{ij}}^l \,\Phi_l &=&  \Phi_i\, \Phi_j 
	&\mathrm{mod} \
        W^\prime \: , \\[0,1cm]
        B\sa( \Psi_\alpha\saa, \Psi_\beta\saa) &=& 
	{{B}_{\alpha\beta}^{\as\,\gamma}}\, \Psi_\gamma\saa
	&=& \Psi_\alpha\saa \Psi_\beta\saa & 
        \mathrm{mod} \left\{ G\sa, 
        \omega^2-p\sa q\sa\right\} \: , \\[0.1cm]  
%
        e\sa(\Phi_i)
         &=& 
        {{e}_{i}^{\as\,\gamma}} \Psi_\gamma\saa
        &=& \Phi_i & \mathrm{mod} \ G\sa \: , \\
        \end{array}
        \end{equation}
where the right-hand side indicates that the equations
hold only modulo the respective relations.
The operator products preserve a ${\bf Z}_2$ grading
defined by $|\Phi_i\saa| = |\Psi_{\ap}\saa| = 0$ and
$|\Psi_{\am}| = 1$, which corresponds to setting
$|\phi| = 0$ and $|\omega| = 1$. Nevertheless we find that in the
present situation, where we have only one chiral superfield in the bulk
action, the boundary structure constants are totally symmetric.
For instance we have $\Psi_{\am}\saa \Psi_{\bm}\saa = \omega \Psit_a 
\omega \Psit_b = \omega \Psit_b\omega \Psit_a = 
\Psi_{\bm}\saa \Psi_{\am}\saa$.

We expect from the experience with the bulk theory that a judicious
``flat'' choice of basis of the chiral ring, as a function of the
deformation parameters, is crucial for the solution of the theory.
Recall how this works in the bulk theory \cite{DVV}: one introduces
suitable polynomials $\Phi_i(\phi,t)$ with the distinguished property
that their 2-point function on the sphere, i.e. the topological
metric, is constant:
  \begin{equation}
  \label{eq:etaij}
  \eta_{ij}\ =\ \langle\,
               \Phi_i(\phi,t)\Phi_j(\phi,t)\,
	            \rangle_{S^2} \ =\ \delta_{i+j,k}
  \end{equation}
(here one made use of $\langle\Phi_{k+2}\rangle_{S^2}=1$). 
The requisite polynomials $\Phi_i(\phi,t)$, including their
dependence on the flat coordinates, can
be explicitly determined in the following simple way \cite{DVV}.
One defines a sequence of $i\times i$ matrices:
 \begin{equation}
	 \label{eq:Cmatrix}
    {C_1^{(i)}(t)}\ = \ 
    {\left(\begin{array}{ccccc}
    0 & 1 & 0 & \cdots&\cdots \\
    t_{2}& 0 & 1 &0 & \\
    t_{3}& t_{2}& 0 & 1& \ddots \\
    \vdots  & \ddots & \ddots & \ddots & \\
    t_{i} & \cdots& t_{3}& t_{2}& 0\\
    \end{array}		  
    \right)}\ ,
 \end{equation}
and in terms of those, one has:
  \begin{eqnarray}
  \label{eq:Wflat}
\Phi_i(\phi,t)\ &=&\ {\rm det}\big(\phi\,\delta_{j}^l-
  {(C_1^{(i)}(t))_j}^l\big)\ \nonumber\\\
    &=& -{\partial\over\partial t_{k+2-i}} \,W_{k+2}(\phi,g(t))
  \ ,\qquad i=0,...,k\ ,\\
    W'_{k+2}(\phi,g(t))\ &=&\ {\rm det}\big(\phi\,\delta_{j}^l-
  {(C_1^{(k+1)}(t))_j}^l\big)\ .\nonumber
  \end{eqnarray}
Note that $C_1^{(k+1)}$ is a matrix representation of the generator
$\Phi_1(\phi,t)$ of the chiral ring, and the last relation corresponds
to the characteristic equation it satisfies. 

We now like to find an analogous polynomial basis for the boundary
ring elements $\{\Psi_\ap\saa,\Psi_\am\saa\} \equiv
\{\Psi_a(\phi,u),\omega\Psit_a(\phi,u)\}$ in the sector $(\Ell)$,
where $u$ are suitable coordinates that need to be determined.

For this we focus on the 2-point function on the disk. Recall that
in the topologically twisted theory, the R-charge has an anomaly
which manifests itself as background charge $q$ in the correlators
\cite{DVV}. For the $A_{k+1}$ minimal models on the disk it takes the
value $q=\hat c = \frac{k}{k+2}$ (as compared to $q=2\hat c$ on
the sphere). From our basis (\ref{eq:basisAk1}) we see that the
only field carrying the correct charge is the top element
$\Psi\saa_{(\Ell,1)}\equiv \omega\Psit_\Ell(\phi,u)$, and it immediately
follows that the overall fermion number of any non-vanishing disk
correlator must be odd.%
\footnote{%
    There occurs a potential subtlety if $k$ is even
    and $\Ell = k/2$, because then both $\Psi_{k/2}$ and
    $\omega\Psit_{k/2}$ have charge $k/(k+2)$ and are potential
    candidates for insertions in non-vanishing correlators.
    However, from a simple analysis with regard to cyclicity properties
    of boundary correlators \cite{LazaroiuTFT} one infers that also in
    this case, $\omega\Psit_{\Ell}$ is the correct top element to consider.
}
More specifically, we can define as the basic non-vanishing correlator
  \begin{equation}
  \label{eq:topcorrel}
  \langle\,\omega\Psit_{\Ell}(\phi,u)\,\rangle\sj_{D^2}\ =\ 1\ ,  
  \end{equation}
and accordingly the metric in the boundary sector $(\Ell)$,
\begin{equation}
  \label{eq:deftopmetric}
  \eta_{\alpha\beta}\sj  = \langle \,
  \Psi_\alpha\saa\Psi_\beta\saa
  \, \rangle_{D^2} \ ,
\end{equation}
must obey:
\begin{equation}
  \label{eq:topmetric}
  \eta_{\alpha\beta}\sj\ =\  
  B^{\as}_{(a,\sigma)(b,\rho)}{}^{(\Ell,1)}\ 
\simeq\ \delta_{\sigma+\rho,1}\ .
\end{equation}
Our aim, thus, is to determine the polynomials $\Psi_a(\phi,u),\Psit_a(\phi,u)$
such that in addition we have:
$B^{\as}_{(a,\sigma)(b,\rho)}{}^{(\Ell,1)}=
\delta_{a+b,\Ell}\ \delta_{\sigma+\rho,1}$.  
This condition does in fact not fix the polynomials uniquely, and
one may impose further physical conditions like integrability of
the correlators.\footnote{Note that in the present context our
notion of flat basis refers to the constancy of the metric, and
not necessarily to the integrability of the correlators.} However
for our present purposes, namely studying the sewing constraints
in the next section, it suffices to invoke the construction outlined
above, and view the ideal in the boundary ring simply as arising
from an auxiliary superpotential:
$G\sj(\phi,u)= {\partial\over\partial\phi}\widehat
W_{\ell+2}(\phi-u_1,u_2,...,u_{\ell+2})$, by writing:
  \begin{equation}
  \label{eq:whatdef}
 G\sj(\phi,h(u))\ =\ {\rm det}\big((\phi-u_1)\,\delta_b^c-
 {(C_1^{(\ell+1)}(u))_b}^c\big)\ ,
  \end{equation}
where $C_1^{(\ell+1)}(u)$ is a matrix just like 
(\ref{eq:Cmatrix}), except that the bulk flat coordinates $t_{k+2-i}$ are 
replaced by the boundary flat coordinates, $u_{\Ell+1-a}$. 
Accordingly, the polynomials
\begin{equation}
       \label{eq:flatPsi}
  \Psit_a(\phi,u)\ =\ \Psi_a(\phi,u)\ =\ 
  {\rm det}\big((\phi-u_1)\,\delta_b^c-
 {(C_1^{(a)}(u))_b}^c\big)\ ,\qquad a=0,...,\ell\ ,
 \end{equation}
satisfy the desired property 
$B^{\as}_{(a,\sigma)(b,\rho)}{}^{(\Ell,1)}=
\delta_{a+b,\Ell}\ \delta_{\sigma+\rho,1}$.
They obey the completeness relation
\begin{equation}
\label{eq:completeness}
\sum_{a=0}^\ell\Psi_{\ell-a}\Psit_a\ =\ \frac{\partial }{\partial x}G\sj(\phi,u)\ ,
\end{equation}
which will prove important later on.

With these ingredients we can obtain an explicit matrix representation
of the boundary chiral ring acting on itself, i.e., of the structure
constants $B_\alpha\sa\equiv{({B}_{\alpha}\sa)_\beta^{\ \gamma}}$
in (\ref{eq:structure}). Concretely, the generators $B_{(1,0)}\cong\phi$
and $B_{(0,1)}\cong\omega$ can be written as:
\begin{eqnarray}
  \label{eq:matrixrep}
  B_{(1,0)}\sa(u) &=&C_1^{(\ell+1)}(u)\otimes
  \left( \begin{array}{cc} 1 & 0 \\ 0 & 1 \end{array}\right)\ ,\\
 B_{(0,1)} \sa(u) &=&   
  \left( \begin{array}{cc} 0 & {\bf 1}\\ 
  p\sa(C_1^{(\ell+1)}(u)+{\bf 1}u_1,v) & 0 \end{array}\right)\ ,
  \end{eqnarray}
and from this the ring relations are easily verified:
\begin{eqnarray}
  \label{eq:matrixvanrel}
  G\sa(B_{(1,0)}\sa(u)) &\equiv& \widehat W'_{\Ell+2}(
  C_1^{(\ell+1)}(u),u)\otimes
  \left( \begin{array}{cc} 1 & 0 \\ 0 & 1 \end{array}\right)\ =\ 0
  \ ,\nonumber\\
  \omega\sa(u,v)\cdot \omega\sa(u,v) &=& 
   p\sa(B\sj_{(1,0)}(u)+{\bf 1}u_1,v)\ .
  \end{eqnarray}

Lowering indices of the ring structure constants,
by contracting with the constant topological metrics 
$\eta_{ij}$ and $\eta_{\alpha\beta}$, these matrices then yield
the boundary preserving
correlators on the sphere and the disk defined by:\footnote
{The bulk-boundary correlators are easily obtained as
follows: $e_{i\,a}\sa =
 (\Phi_i(B_{(1,0)}\sa(u),t)\cdot \Psit_a(B_{(1,0)}\sa(u),u))_1^{\ 2\ell+2}$.}
\begin{eqnarray}
  \label{eq:diskcorrs}
  C_{ijl}(t(u,v)) &=& \langle\,\Phi_i\Phi_j\Phi_l\,\rangle_{S^2} ,
  \nonumber\\
  B_{\alpha\beta\gamma}\sa(u,v) &=& \langle\,\Psi_\alpha\saa  
  \Psi\saa_\beta\Psi\saa_\gamma\,
  \rangle_{D^2}\sa ,\\
    e_{i\alpha}\sa(u,v) &=& \langle\,\Phi_i\Psi\saa_\alpha\,
    \rangle_{D^2}\sa . \nonumber
  \end{eqnarray}
%
Here, taking $\langle...\rangle_{S^2}$ amounts to extracting the
part proportional to $\Phi_k$ and $\langle...\rangle_{D^2}\sj$
amounts to extracting the part proportional to $\omega\Psit_{\Ell}$,
modulo the relevant relations in the ring. Note that
in the present situation with one LG variable,
both bulk and boundary 3-point functions
are symmetric in the indices. 

With these building blocks, we can supposedly
determine any correlator in the theory in a fixed given boundary
sector $(\Ell)$. However, despite having constructed some
flat basis of the ring of physical operators and having expressed
the sphere and disk amplitudes in terms of them, it is not yet
clear whether the objects (\ref{eq:diskcorrs}) really define a
consistent open string TFT - what remains to be done is to verify
that the topological sewing constraints are indeed satisfied.

Before we will do so in the next section, let us recall that the
open string sewing constraints are the defining axioms of an (in
general non-commutative) extended Frobenius manifold
\cite{Natanzon}.  So if they are satisfied, and this is what we will show
below, we know from the general results of \cite{Natanzon}\ that there exists
a formal ``structure series'' whose derivatives generate the disk
correlators of the theory. In a string theory context, this
topological disk partition function would correspond to the
effective $N=1$ superpotential on the brane world-volume.

A more detailed discussion of flat coordinates in relation with the
integrability of the correlators is beyond the scope of our present
paper, and we defer it to forthcoming work in preparation.

\subsection{Open-closed sewing constraints in LG formulation}

We will now verify that the family of generically perturbed boundary
LG models with one variable indeed obeys the sewing constraints
of an open-closed TFT. As in the preceding section, we restrict
ourselves to the algebra of boundary preserving fields.
In \cite{LazaroiuTFT,MooreSegal} the open-closed TFT was axiomatically
formulated in terms of five sewing constraints. Two of those
correspond to the associativity of the bulk and boundary operator
products, respectively. Two further bulk-boundary crossing relations
control the behavior of bulk fields when moving to the boundary,
and finally the generalized topological Cardy condition serves as
a link between the open and the closed string sectors.

Since both the bulk ring (\ref{eq:bulk ring}) and
the boundary ring (\ref{eq:boundary ring}) of chiral primary fields
are defined by polynomial multiplication modulo some fixed polynomials,
the pure bulk and boundary crossing symmetry relations are satisfied
by construction. Moreover, the bulk-boundary crossing 
relation \cite{LazaroiuTFT}%
\footnote{Here we can omit the sign factor which occurs in the general
          formulation of this constraint because the bulk fields are
          all bosonic. 
}%
        \begin{equation}
        \label{eq:1stbubo}
	B\sj \bigl(\, \Psi\saa_\alpha,\, e\sj(\Phi_i) \,   \bigr) = 
	B\sj \bigl(\, e\sj(\Phi_i)   ,\, \Psi\saa_\alpha \,\bigr)
        \end{equation}
is trivially satisfied, since in our situation with one
bulk LG variable the boundary structure constants are totally
symmetric for boundary preserving operators. 

The only non-trivial bulk-boundary crossing symmetry is
        \begin{equation}
        \label{eq:2ndbubo}
	B\sj \bigl(\, e\sj(\Phi_i),\, e\sj(\Phi_j) \,\bigr) =
	e\sj\bigl(\,C(\Phi_i,\Phi_j) \,\bigr) \;.
        \end{equation}
This equation means that $e\sj$ is a morphism from the bulk to the
boundary ring and therefore gives rise to a relation between the
polynomials $W'(\phi)$ and $G\sa (\phi)$.
We will show that this connection is indeed already implied by the
factorization condition (\ref{eq:WEJ}) together with (\ref{eq:EJfact}). 

Let us take two general bulk fields $P_i$ and $P_j$ of polynomial
degree $i$ and $j$.
If we plug these polynomials into relation (\ref{eq:2ndbubo}), we
have to distinguish between the cases $r < k+1$ and $r \geq k+1$,
where $r = i+j$. In the first case we get the trivial statement
	\begin{displaymath}
	\left[ P_i \: \mathrm{mod} \: G \sa (\phi) \right] \cdot
	\left[ P_j \: \mathrm{mod} \: G \sa (\phi) \right] =
	P_i P_j \: \mathrm{mod} \: G \sa (\phi) \: , 
	\end{displaymath}
because we need not make use of the vanishing relation
$W^\prime(\phi) = 0$. As for the second case, if we write
$P_i P_j = r(\phi) W^\prime(\phi) + s(\phi)$, the constraint boils down to:
        \begin{equation}
        \label{eq:2ndbuboeqn1}
	\begin{array}{ccc}
        r(\phi) W^\prime(\phi) & \mathrm{mod} & G \sa (\phi) = 0 \: .
	\end{array} 
        \end{equation}
Choosing $r = k+1$ we have $r(\phi) = 1$ and the above   
condition becomes
        \begin{equation}
        \label{eq:2ndbuboeqn2}
	\begin{array}{ccc}
        W^\prime(\phi) & \mathrm{mod} & G \sa (\phi) = 0  \: .
	\end{array}
        \end{equation}
For $r > k+1$ condition (\ref{eq:2ndbuboeqn2}) is already sufficient
for (\ref{eq:2ndbuboeqn1}) to be satisfied.

On the other hand, if we use the factorization condition
(\ref{eq:WEJ}) as well as the decompositions (\ref{eq:EJfact}), we can
write the superpotential $W(\phi)$ as
        \begin{displaymath}
        W(\phi) = p \sa (\phi) q \sa (\phi) \left[ G \sa (\phi)
	\right]^2 + {\rm const.} \;. 
        \end{displaymath}
The derivative $W^\prime(\phi)$ is then
        \begin{displaymath}
        W^\prime(\phi) = \left\{ \left[ p \sa (\phi) q \sa (\phi)
	\right]^\prime G \sa (\phi) + 2 p \sa (\phi) q \sa (\phi)
	G^{\as \prime}(\phi) \right\} G \sa (\phi) \: ,
        \end{displaymath}
which implies (\ref{eq:2ndbuboeqn2}). 
We thus see that the topological sewing constraint
(\ref{eq:2ndbubo}) is already satisfied as a consequence of the
conditions for a supersymmetric action. 

The remaining consistency condition that needs to be checked
is the topological Cardy condition \cite{LazaroiuTFT,MooreSegal}. 
%
%
In order to formulate it conveniently, we
first introduce the adjoint boundary-bulk mapping, $f\sj$. It is defined
by \cite{LazaroiuTFT} 
%
%
%
\begin{equation}
  \label{eq:defadjoint}
  \langle e\sj(\Phi_i) \; \Psi\saa_\alpha 
  \rangle\sj_{D^2}  = 
  \langle \Phi_i \; f\sj(\Psi\saa_\alpha) 
  \rangle_{S^2}\; ,
\end{equation}
and in our LG theory it takes the form: 
\begin{eqnarray}
  \label{eq:adjoint}
  f\sj(\omega \Psit_a(\phi)) &=& 
  \{(p\sj q\sj)'G\sj + 2 p\sj q\sj G\sj{}'\} \; \Psit_a(\phi)\;,\\
  f\sj(\Psi_a(\phi) )        &=& 0 \;.
\nonumber
\end{eqnarray}
Note that $f\sj$ is consistent with the truncations of the bulk and
the boundary ring, i.e. $f\sj(\omega G\sj) = W'$. The second
expression in (\ref{eq:adjoint}) vanishes identically because of the
fermionic character of the boundary metric.  

\begin{figure}[t]  \label{fig1}
\begin{center}
\epsfxsize=14cm
\epsffile{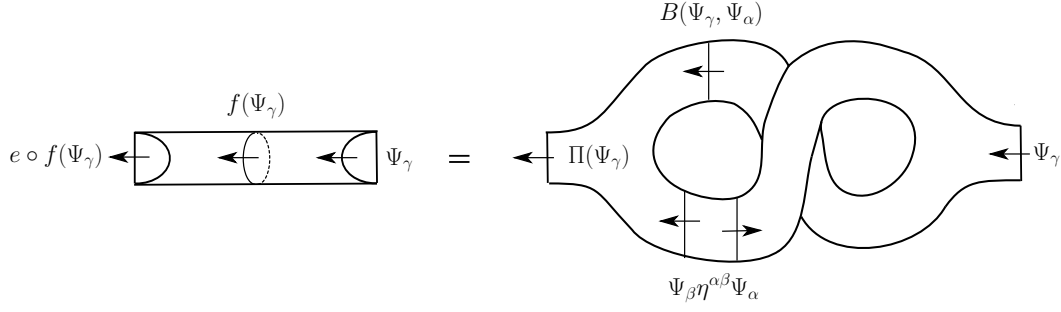}
\caption{The Cardy condition requires that the factorization
of the topological annulus amplitude on closed and open string channels
yields the same result. We show in the text that its solution is intimately tied
to the factorization condition $W=EJ$ of the LG superpotential.}
\label{fig:Cardy}
\end{center}
\end{figure}

We are now prepared to formulate the topological Cardy condition and 
to describe how it is satisfied in the Landau-Ginzburg
theory. The Cardy constraint, which we write in the form:
\begin{equation}
  \label{eq:Cardytwist}
  e\sj \circ f\sj (\Psi\saa_\gamma)= 
  \Pi\sj (\Psi\saa_\gamma)\;,
\end{equation}
relates the two ways the topological annulus amplitude depicted in figure
\ref{fig:Cardy} can be decomposed into open and closed string channels.
The left-hand side of (\ref{eq:Cardytwist}) corresponds to the closed
string channel, whereas the right-hand side is the double-twist
diagram,
\begin{equation}
  \label{eq:double twist}
  \Pi\sa(\Psi\saa_\gamma) \ \equiv\  \sum_{\alpha,\beta} 
	      (-)^{(|\gamma|+|\alpha|)|\alpha|} \; \eta\sj{}^{\beta\alpha}
              B\sj\bigl(\,\Psi\saa_\beta,\,B\sj(\Psi\saa_\gamma,\,
	      \Psi\saa_\alpha)\,\bigr)  \;,
\end{equation}
of the open string channel. The sign in
(\ref{eq:double twist}) comes from the twist on the open string side of
figure \ref{fig:Cardy}. Using (\ref{eq:adjoint}), the left-hand side
of (\ref{eq:Cardytwist}) becomes 
\begin{equation}
  \label{eq:right Cardy}
  e\sj \circ f\sj (\omega\Psit_a(\phi)) = 
  (2\,p\sj q\sj\,G\sj{}')\,
  \Psit_a(\phi) \quad\textrm{mod}\quad G\sj\;.
\end{equation}
We remember that the $f$-mapping of a bosonic insertion vanishes
trivially. 
In order to evaluate the double-twist side, we use the basis
$\{\Psi_a,\omega\Psit_b\}$ with the off-diagonal metric
$\eta_{ab}=\delta_{a+b,\Ell}$. For bosonic fields the double-twist diagram
leads to zero, as it should be, because bosonic and fermionic
contributions in (\ref{eq:double twist}) cancel each other, i.e.,  
\begin{eqnarray}
  \Pi\sj(\Psi_c(\phi)) &=& 
  \sum_{a,b} 
  \bigl( \;\eta^{ba} 
         B\sj(\omega\Psit_b,B\sj(\Psi_c,\Psi_a)) \nonumber\\
& &\;\;-\; \eta^{ba} 
         B\sj(\Psi_b,B\sj(\Psi_c,\omega\Psit_a))
  \bigr) = 0 \;.
\end{eqnarray}
Applied to a fermionic field, equation (\ref{eq:double twist}) becomes
\begin{eqnarray}
  \Pi\sj(\omega \Psit_c(\phi)) &=& 
  \sum_{a,b} 
  \bigl( \eta^{ba} 
         B\sj(\omega\Psit_b,B\sj(\omega \Psit_c,\Psi_a)) +
	 \eta^{ba} 
         B\sj(\Psi_b,B\sj(\omega \Psit_c,\omega\Psit_a))
  \bigr) \nonumber\\
  &=& 2 \;\omega^2\,\Psit_c
      \sum_{a,b} \eta^{ba}\Psi_a\Psit_b
      \quad\textrm{mod}\quad G\sj\;,
\end{eqnarray}
and using relation
$\sum_{a,b} \eta^{ab}\Psi_a(\phi)\Psit_b(\phi) = G\sj{}'$ (\ref{eq:completeness})
for the flat basis, we obtain
\begin{equation}
  \label{eq:left Cardy}
  \Pi\sj(\omega \Psit_c) = 
  (2\,\omega^2\,G\sj{}')\,\Psit_c 
  \quad\textrm{mod}\quad G\sj\;.
\end{equation}
The comparison of (\ref{eq:right Cardy}) and (\ref{eq:left Cardy})
shows that the fermionic ring relation (\ref{eq:omega}),
$\omega^2=p\sj q\sj$, is a crucial ingredient in order to satisfy the 
Cardy relation. The other important ingredient, the factorization
$W = p\sj q\sj G\sj{}^2 +$ const., enters the Cardy condition
through the adjoint mapping (\ref{eq:adjoint}).

Before closing this section we want to make a remark on
topological sewing constraints for boundary changing operators, since
there occur some subtleties. The topological metric
(\ref{eq:deftopmetric}) for boundary changing operators 
(\ref{eq:changing basis}) is generically degenerate. Therefore, the
Cardy condition cannot be formulated in terms of the double-twist diagram
(\ref{eq:double twist}), because it contains explicitly the inverse
metric. Moreover, one can show that the bulk boundary crossing relation
(\ref{eq:1stbubo}) is only satisfied in the sense of Ward identities,
i.e., only in correlation functions and not as operator identities. 
This might suggest a relaxation of some of the axioms \cite{LazaroiuTFT} for a
topological field theory.

\section{Categorial description of $B$-type $D$-branes}

We know from \cite{Kontsevich,Douglas}\ 
that $D$-branes can often be mathematically described in
the language of categories. The branes, or equivalently
the boundary conditions or boundary states, provide the objects
of the category, whereas the open strings stretching between
the $D$-branes are the morphisms. 
Direct contact between a mathematical description of this
type and a field theoretical approach has been made in \cite{Douglas,AsLa}
 in the context
of B-type branes on Calabi-Yau manifolds, where it was shown
that the derived category of coherent sheaves on a manifold $X$
can be obtained as a category of boundary conditions in the
B-type topologically twisted sigma model on $X$.
To generalize these ideas to Landau-Ginzburg models, the
essential new ingredient that has to be taken into account
is the superpotential $W$, or, in other words,
a regular function $W$ on $X$. 

A mathematical definition of B-branes for such models
has been proposed by Kontsevich, as reviewed in
\cite{orlov},%
\footnote{In that paper it was shown that the B-type branes 
can be equivalently described in terms of a category
$D_{Sg}(X)$, which is tied more closely to the singularity structure
of $X$ rather than to the ``rest'' of it.} and 
investigated in a physics context in \cite{KapustinLG}. 
In this section, we will work out this description of
$D$-branes for the open string TFT with superpotential 
$W=\frac1{k+2}\phi^{k+2} + \ldots$, where the dots denote general
perturbations; we will find
that our results of the previous sections are equivalent to
Kontsevich's in that the underlying relevant cohomologies
are isomorphic. 

Let us briefly recapitulate the construction of Kontsevich,
where we closely follow \cite{orlov}.
The first step is to define a triangulated category $DB_{w}$
for each value $w \in \BC$. The category of B-type
branes is then obtained as the disjoint union of all $DB_w$.
As shown in \cite{orlov}, only finitely many $w$ contribute
to this union, namely only those corresponding to critical
points of the superpotential.
For our purposes, the relevant variety will be $X=\BC$, which simplifies
the general discussion in \cite{orlov}, and
the only relevant value for $w$ is $w=0$.
The category $DB_0$ is then defined in the following way:

\subsubsection*{The objects}

The objects of the category are ordered pairs
\begin{equation}
\overline{P}:=\Bigl(
\xymatrix{
P_1 \ar@<0.6ex>[r]^{p_1} &P_0 \ar@<0.6ex>[l]^{p_0}
}
\Bigl) \;.
\end{equation}
In the general case, $P_0$ and $P_1$ are projective $A$-modules, where
$A$ is such that the smooth variety $X$ is obtained as $X=Spec(A)$.
In the simple case that we consider,
where $X$ is the complex plane, the
only relevant projective module corresponds to the structure
sheaf $\cO$, so that $P_0=P_1=\cO$. 
The choice for the maps $p_0$, $p_1$ is restricted by the 
requirement that their
composition is the multiplication by $W$. As we will see,
they correspond to the polynomials $E$ and $J$ in (\ref{eq:WEJ}).

\subsubsection*{The morphisms}

The morphisms of the category are given by
\begin{equation}
Hom(\overline{P},\overline{Q}) = \bigoplus_{i,j} Hom(P_i,Q_j),
\end{equation}
subject to the restriction that they are closed with respect
to a differential operator $D$, and taken modulo 
$D$-exact operators.
We define the differential $D$ acting on a morphism 
$f \in Hom(P_i,Q_j)$ by
\begin{equation}
D f=q \circ f-(-1)^{k}f\circ p .
\end{equation}
Here, $k$ is a $\BZ_2$ grade of $f$,  
given by $i-j$. We refer to degree $0$ operators as bosons and
to degree $1$ operators as fermions.
The differential maps even to odd morphisms,
i.e. is itself an odd operator, as it should be.
 
To determine the open string spectrum on a $D$-brane in the language
of categories, we will now spell
out explicitly the conditions for an operator to
be physical.%
\footnote{Note that our discussion
differs slightly from the one given in \cite{orlov}: 
we accept $D$-closed morphisms of both even and odd
degree as physical operators, whereas \cite{orlov} imposes
a further restriction
to operators of even degree.}
A bosonic operator $f$, which
maps $\overline{P}\to \overline{Q}$,
consists of two components
$f=(f_0,f_1)$, where $f_0: P_0\to Q_0$ and $f_1: P_1\to Q_1$. 
The differential $D$ acts as
\begin{equation}
Df= \left( \begin{array}{c} q_0 f_0-f_1 p_0 \\ q_1 f_1 -f_0 p_1 \end{array}
\right)\ ,
\end{equation}
and this implies in particular that the condition
$Df=0$ can be formulated in terms of the components as 
\begin{equation} \label{bosexact}
q_1 f_1 =
f_0 p_1, \quad  q_0 f_0 = f_1 p_0.
\end{equation}
Likewise, a fermionic operator $t$
has two components, $t=(t_0,t_1)$, where $t_0:P_0\to Q_1$ and
$t_1:P_1 \to Q_0$. The differential acts on the fermions as
\begin{equation}
Dt = \left( \begin{array}{c} q_1 t_0+t_1p_0 \\ q_0 t_1 + t_0 p_1 \end{array}
\right)\ .
\end{equation}
For the bosonic spectrum, we thus want to divide out the operators
that can be written as 
\begin{eqnarray}\nonumber
f_0 &=& q_1 t_0 +t_1 p_0 \ ,\\ 
f_1 &=& q_0 t_1 +t_0 p_1.
\end{eqnarray}
The conditions for a fermionic operator $(s_0,s_1)$
to be in the physical spectrum
are
\begin{equation}
q_1 s_0 = -s_1 p_0, \quad q_0 s_1 = - s_0 p_1,
\end{equation}
modulo the operators that are derivatives of a  boson $(g_0,g_1)$
\begin{eqnarray}\nonumber
s_0 &=& q_0 g_0 - g_1 q_0 \ ,\\ \label{fermexact}
s_1 &=& q_1 g_1 - g_0 p_1 \ .
\end{eqnarray}

It is sometimes useful to summarize  the operators in
matrix notation in the following way:
\begin{equation}
{\cal F}_{\overline{P}\overline{Q}} = \left( 
\begin{array}{cc} f_0 & 0 \\ 0 & f_1 \end{array}
\right), \quad
{\cal S}_{\overline{P}\overline{Q}} = \left( 
\begin{array}{cc} 0 & s_1 \\ s_0 & 0 \end{array}
\right)
\ .
\end{equation}
Here, ${\cal F}$ and ${\cal S}$ are bosonic and fermionic operators,
respectively, and it is understood that $f_0$ maps $P_0$ to $Q_0$, $f_1:
P_1 \to Q_1$, $ s_0: P_0 \to Q_1$ and $s_1: P_1 \to Q_0$.
The matrix multiplication is then compatible with the
composition of operators. It is possible to represent
also the derivative $D$ in
terms of matrices. For this, define
\begin{equation}
{\cal Q} = \left( \begin{array}{cc} 0 & q_1 \\ q_0 & 0 \end{array} \right),\quad
{\cal P} = \left( \begin{array}{cc} 0 & p_1 \\ p_0 & 0 \end{array} \right), \quad
\epsilon =  \left( \begin{array}{cc} 1 & 0 \\ 0 & -1 \end{array} \right).
\end{equation}
The derivative acting on a matrix ${\cal F}$ can then be expressed as
\begin{equation}
D {\cal F}_{\overline{P} \overline{Q}} =
{\cal Q} {\cal F}_{\overline{P} \overline{Q}} - 
\epsilon {\cal F}_{\overline{P} \overline{Q}} \epsilon {\cal P}\ .
\end{equation}
Since the category $DB_0$ is triangulated, there exists a translation
functor denoted by ``$[1]$'', or, in physics language, 
a notion of an anti-brane. It is defined by
\begin{equation}
\overline{P}[1]=\xymatrix{
\Bigl(
P_0 \ar@<0.6ex>[r]^{-p_0} &P_1 \ar@<0.6ex>[l]^{-p_1}
}
\Bigl).
\end{equation}
The spectrum of bosonic physical operators between
$\overline{P}$ and $\overline{Q}[1]$ coincides with the fermionic
part of the spectrum between $\overline{P}$ and $\overline{Q}$.
Switching to anti-branes shifts the grade of all operators by one unit.

To give the full data of a triangulated category, we have to
define a set of standard triangles in the category. To do so,
we first associate to any morphism $f:\overline{P}\to\overline{Q}$ 
a mapping cone $C(f)$ as an object
\begin{equation}
C(f)=\xymatrix{
\Bigl(
Q_1\oplus P_0 \ar@<0.6ex>[r]^{c_1} & Q_0\oplus P_1 \ar@<0.6ex>[l]^{c_0}
}
\Bigl)\ ,
\end{equation}
such that
\begin{equation}
c_0=
\begin{pmatrix}
q_0 & f_1\\
0 & -p_1
\end{pmatrix},
\qquad
c_1=
\begin{pmatrix}
q_1 & f_0\\
0 & -p_0
\end{pmatrix}.
\end{equation}
Then there are  maps
$g: \overline{Q}\to C(f), \; g=(\id , 0)$ and $h: C(f)\to\overline{P}[1],\;
h=(0, -\id)$, and the standard triangles are given as
\begin{equation} \label{triangle}
\overline{P}\stackrel{f}{\lto} \overline{Q}\stackrel{g}{\lto} C(f)
\stackrel{h}{\lto} \overline{P}[1].
\end{equation}

To make the connection to physics, note that the triangles
are the appropriate language to discuss tachyon condensation
\cite{AsLa}: The tachyon corresponds to the map $f$, representing
an open string state. ``Tachyon condensation'' means to form the
``sum'' of two branes $\overline{P}$ and $\overline{Q}$ and to
deform by $f$. The result is a single $D$-brane, mathematically described by the
cone, $C(f)$. The meaning of the triangle (\ref{triangle})
is that $\overline{P}$ and $\overline{Q}$ can combine to give
$C(f)$ after tachyon condensation.

\subsubsection*{Calculation of the spectrum}

As already mentioned, for an arbitrary Landau-Ginzburg model in one variable, 
the only relevant projective module to consider is $\cO$. The maps $p_0$
and $p_1$ are polynomials whose product is $W$. On a single $D$-brane
the derivative $D$ 
acts on operators that are either purely bosonic or
purely fermionic as $D {\cal F} = [ {\cal P}, {\cal F}]_{\pm}$,
where, as usual, one picks the commutator if ${\cal F}$
is bosonic and the anti-commutator if ${\cal F}$ is fermionic.

The condition for $D$-closedness for bosons on a single brane is simply 
$f_0=f_1$, so that the bosonic physical operators are
diagonal matrices. The matrix multiplication of two bosons reduces to the
multiplication of holomorphic polynomials $f_0$ in one variable $z$. 
Polynomials of the type $f_0 = t_0 p_1 + t_1 p_0$, where $t_0, t_1$
are arbitrary, are divided out. To solve for the cohomology, let
us decompose $p_0$ and $p_1$ as
\begin{equation}
p_0 = G^{p_0p_1} \hat{p}_0^{p_1}, \quad \quad p_1= G^{p_0p_1} \hat{p}_1^{p_0},
\end{equation}
where $G^{p_0p_1}$ is the greatest common divisor of $p_0$, $p_1$.
One can then see that $f_0$ has to be taken modulo $G^{p_0p_1}$.

For the fermions, $D$-closedness means that $p_1 s_0=-p_0 s_1$, where,
according to (\ref{fermexact}), $s_0$ has to be taken modulo $p_0$,
and $s_1$ is defined modulo $p_1$. It follows immediately that if
one has two  fermionic solutions to these equations, they can only
differ by multiplication by a diagonal matrix. Hence, all physical
fermions are of the form $\omega p(z)$, where $\omega$ is a solution
to the constraint equation for the fermions and $p(z)$ a polynomial in $z$
corresponding to a physical boson.
We can thus write the following expression for $\omega$
\begin{equation}
\label{eq:omega matrix}
\omega = \alpha \left( \begin{array}{cc} 0 & -\hat{p}_1^{p_0} \\ 
\hat{p}_0^{p_1} & 0
\end{array} \right)
\end{equation}

Notice that the computation of the cohomology we have just outlined is strongly reminiscent
of the computation in the boundary LG theory,
presented in section 3.2.
To show that these computations are in fact isomorphic, observe that
\begin{equation}
D \left( \begin{array}{cc} 0 & 0 \\ 1 & 0 \end{array} \right)
= p_1, \quad
D \left( \begin{array}{cc} 0 & 1 \\ 0 & 0 \end{array} \right)
= p_0,
\end{equation}
which reproduces (\ref{eq:Qvar}) if we identify 
\begin{equation}
{\cal P} = \sqrt{i} Q, \quad
\pi =  \left( \begin{array}{cc} 0 & 0 \\ \sqrt{i} & 0 \end{array} \right), \quad
\bar{\pi} = \left( \begin{array}{cc} 0 & -i \sqrt{i} \\ 0 & 0 \end{array} \right),
\quad p_0=J, \quad p_1 = E.
\end{equation}
If we set $\alpha = i$ in (\ref{eq:omega matrix}) and use the above
identification we get back the expression (\ref{eq:omegadef}).
This shows
explicitly that the cohomology problem of the Kontsevich approach
is exactly the same as
the one encountered in the Landau-Ginzburg formulation. Therefore,
the spectrum necessarily agrees in the two formulations.

The same holds for the boundary changing operators: since at
this point we can  map the cohomology problem
to the equivalent problem in the Lagrangian
approach, we omit an explicit analysis of those operators
in the language of categories.

To recover the structure of the boundary rings discussed in earlier sections,
note that ``taking the o.p.e.'' corresponds to the composition
of morphisms. In this way, the Kontsevich approach reproduces the
boundary structure constants in the second line of (\ref{eq:structure}).

\subsubsection*{The restriction to $W=z^{k+2}$}

Although it is clear from the above arguments that the spectrum
of boundary preserving and boundary changing operators
for the special case $W=z^{k+2}$ agrees exactly with
the one obtained from the LG theory, we find it
an instructive exercise to explicitly work out the full
spectrum for this simple case.
To specify boundary conditions,
we choose $p_1= z^{\mu}$, which determines $p_0= z^{k+2-\mu}$.
The bosonic open string spectrum between the brane $\overline{P}$ with
$(p_0,p_1)=(z^{k+2-\mu}, z^{\mu})$ and the brane $\overline{Q}$ with
$(q_0,q_1)= (z^{k+2-\nu}, z^{\nu})$ 
is determined using (\ref{bosexact}), which becomes
$$
f_1 z^{k+2-\mu} = f_0 z^{k+2-\nu}, \quad f_0 z^\mu = z^\nu f_1,
$$
which is to be taken modulo 
\begin{eqnarray*}
f_0 &=& t_0 z^{k+2-\mu} + t_1 z^{\nu} \\
f_1 &=& t_0 z^{k+2-\nu} + t_1 z^{\mu}\ .
\end{eqnarray*}
Evaluating these conditions, we conclude that for $\mu\geq\nu$
the physical operators
are
\begin{equation}
\left( \begin{array}{cc} f_0 & 0  \\ 0 & f_1 \end{array} \right)_l
= \left( \begin{array}{cc} z^l & 0\\ 0 & z^{l+\mu-\nu} \end{array} \right)\ .
\end{equation}
Here, $l$ can take the values $l= 0,\dots, \min\{\nu, k+2-\mu\} -1
= \min \{ \nu, \mu, k+2-\nu, k+2-\mu \}-1$.
Similarly, for $\mu \leq \nu$ we get
\begin{equation}
\left( \begin{array}{cc} f_0 & 0 \\ 0 & f_1 \end{array} \right)_l
= \left( \begin{array}{cc} z^{l-\mu+\nu} & 0  \\ 0 & z^{l} \end{array} \right),
\end{equation}
where $l$ can take the values $0,\dots, \min \{ \mu, k+2 -\nu \}-1
= \min \{ \nu, \mu, k+2-\nu, k+2-\mu \} -1$.
A similar analysis for the fermions leads to the following
spectrum of operators:
\begin{equation}
\left( \begin{array}{cc} 0 & s_1 \\ s_0 & 0 \end{array} \right)_l
= \left( \begin{array}{cc} 0 & -z^{l+\mu +\nu -k-2} \\ z^{l} & 0
\end{array} \right),
\quad l=0, \dots, \min \{ k+2-\mu, k+2 -\nu \}-1 \;,
\end{equation}
for $\mu +\nu \geq k+2$ or
\begin{equation}
\left( \begin{array}{cc} 0 & s_1 \\ s_0 & 0 \end{array} \right)_l
= \left( \begin{array}{cc} 0 & z^{l} \\ -z^{l+k+2-(\mu +\nu)} & 0 \end{array} \right),
\quad l=0, \dots, \min \{ \mu, \nu \} -1 \;,
\end{equation}
for $\mu +\nu \leq k+2$. 
The spectrum obtained in this way agrees perfectly with
the one obtained from the boundary Landau-Ginzburg model (setting 
$\mu=\Ell_1\!+\!1$ and $\nu=\Ell_2\!+\!1$),
as well as with the boundary conformal 
field theory results summarized in Appendix~A below.

\vfill\break
\appendix

\section*{Appendix A: Boundary spectrum of minimal models from CFT}
\label{sec:CFT}
\renewcommand{\theequation}{A.\arabic{equation}}
\setcounter{equation}{0} 

The $N=2$ minimal model can be  realized as an $SU(2)$ WZW model
and a Dirac fermion, coupled through a $U(1)$ gauge field. The
symmetry group is $\BZ_{2k+4} \times \BZ_2$, where $\BZ_{2k+4}$
is an axial $R$-rotation whose generator is denoted by $a$ and $\BZ_2$
is the fermion number $(-1)^F$.%
\footnote{More precisely, $a= e^{\pi i J_0}$, where $J_0$ is
          the zero-mode of the $U(1)$ $R$-current.
}
Taking the orbifold by $(-1)^F$ (a non-chiral GSO-projection)
one obtains the rational conformal field theory 
$SU(2)_k\times U(1)_2/U(1)_{k+2}$.
Its $D$-branes can be studied using standard BCFT techniques; their relation
to geometry has been studied in \cite{HIV,ADEALE,Maldacena:2001ky}.

In order to compare with the results of the present paper obtained from the LG model, 
we are interested to obtain the spectrum 
on B-type $D$-branes in the $unprojected$ theory,
including the statistics of the boundary operators. 
Starting from the B-type boundary states of the rational
model, one first has to undo the GSO projection to obtain the
boundary states in the unprojected theory. One can then identify
the action of $a$ and $(-1)^F$ in the open string sector;
the latter in particular determines the statistics.
These steps have been performed  in \cite{BH}, and we
refer to that paper for a detailed discussion. For completeness,
we summarize the main steps and the result.

The primary fields of the rational model are labeled by the triple
$(l,m,s)$ where $l \in \{ 0,1,2,...,k \}$, $m$ 
is an integer modulo $2k+4$, and $s$ is an integer modulo 4. The NS
sectors are defined by $s=0,2$ and the R sectors by $s=-1,1$. We also
have the identification $(l,m,s) \sim (k-l,m+k+2,s+2)$ and the
selection rule $l + m + s = 0$ mod 2. The chiral primary (antichiral
primary) states in the NS sector are labeled by $(l,l,0$)
($(l,-l,0)$) if we use the identification in order to set $s=0$.
The symmetry group of the model is $\BZ_{4k+8}$ 
(generated by the simple current $(0,1,1)$) for $k$ odd
and $\BZ_{2k+4}\times \BZ_2$ (generated by $(0,1,1)$ and 
$(0,0,2)$) for $k$ even. The current $(0,0,2)$ distinguishes
the R and NS sectors of the theory and can be viewed as the
quantum symmetry of $(-1)^F$. 

The  Cardy states (A-type boundary states) $|L,M,S \rangle_C$ 
are labeled by the same set $(L,M,S)$ as the primary states. 
B-type boundary states can
be constructed using the fact that one can obtain the diagonal 
form of the charge conjugation modular invariant by taking a $\BZ_{k+2}
\times \BZ_2$ orbifold. Hence, taking $\BZ_{k+2}
\times \BZ_2$ orbits of A-type states plus an application of the
``mirror map'' (charge conjugation on the left-movers) leads to
B-type boundary states. The $\BZ_{k+2}$ acts on the Cardy states by
shifting $M$ by $2$ and the $\BZ_2$ acts by shifting $S$ by $2$.
We therefore label B-type
states by the orbit labels
$L = \{0,1,2,...,[\frac{k}{2}]\}$, $M = 0$ and $S =
0,1$. All of these states are purely in the NSNS sector, and these
branes are unoriented. A special
case arises  for the case $k$ even and $L = \frac{k}{2}$ (this observation
traces back to  \cite{Sagnotti}). In this
case the orbit boundary state is not elementary but can be
decomposed further: 
There are altogether four states
$|B,\frac{k}{2},\hat{S} \rangle$ with $\hat{S} = -1,0,1,2$,
which are linear combination of an ``orbit'' NSNS part 
$|B,\frac{k}{2},S\rangle$
(where $S$ is the mod $2$ reduction of $\hat{S}$) and an extra
RR piece. In particular, these branes are oriented. 
We refer to \cite{Maldacena:2001ky}\ for details of the construction.

The task is now to resolve the GSO projection to obtain the branes
of the unprojected theory. As explained in \cite{BH}, the
unoriented branes remain the same in the projected and unprojected
theory. On the other hand, the oriented (short orbit) branes
get re-decomposed into a NSNS  and an RR part. In this paper, we
have developed a LG formulation of  the unoriented orbit-type branes, and we
point out that a LG interpretation of the oriented B-type branes has been proposed
by the authors of \cite{mirbook}. 

The open string spectrum between the unoriented branes can
be obtained as 
\begin{equation}
  \label{eq:spec}
	{\cal H}_{(L,S)(L',S')} = 
	\bigoplus_{l+m+s \, {\rm even}}
	N_{LL'}^{l} \Scr{H}^{N=2}_{l,m,S-S'},
\end{equation}
where $N_{LL'}^l$ are the $SU(2)_k$ fusion rule coefficients.
The spaces ${\Scr H}^{N=2}_{l,m,[s]}$ are the modules of the 
unprojected $N=2$
theory, which can be written in terms of the GSO-projected
modules as ${\Scr H}^{N=2}_{l,m,[s]}={\Scr H}_{l,m,s}
+{\Scr H}_{l,m,s+2} $. $[s]$ denotes the mod $2$ reduction of $s$
and distinguishes NS and R sectors. (Note that $S$ and $S'$ in
(\ref{eq:spec}) were only defined mod $2$, therefore 
$[S-S'] = S-S'$ and the bracket can be omitted.)

Since these boundary states are purely in the NSNS sector, it is
clear from the closed string sector that the
Witten index between them vanishes. For the R-ground states in
the open string sector this means that their contributions to
${\rm tr} (-1)^F$ cancel out, in other words, half of the supersymmetric
$R$ ground states are bosonic, and half of them are fermionic.
More precisely, one can see that on a $(L,S)(L',S\!+\!1)$-brane pair 
the ground states from ${\Scr H}^{N=2}_{l,l+1,1}$ and 
${\Scr H}^{N=2}_{l,-l-1,1}$ (which is an element of
the Hilbert space ${\Scr H}_{l,-l-1,-1}$ of the GSO-projected theory)
contribute with opposite sign \cite{BH}.

By spectral flow $(0,\!-\!1,\!-\!1)$ these representations are related to 
${\Scr H}^{N=2}_{l,l,0}$ . Note however that the spectral flow operator
is not part of the spectrum of a single brane: RR ground states only
propagate if $S-S'=1$ mod $2$ and NSNS states only if $S-S'=0$ mod $2$.
In particular, there are never RR states on a single brane.
It is natural to assume that the NSNS chiral primaries split
up into a set of bosonic and fermions just as their RR counter parts,
which propagate between branes with appropriately shifted label $S$. 

To be explicit, the chiral ring consists of elements with charges
($\tilde q = q\,(k\!+\!2)$)
\begin{eqnarray}
  \label{eq:BCFT ring}
  \begin{array}{l@{\;\;\in\;\;}l@{\quad\mathrm{in}\;\;}l}
  \tilde q = l & \left\{ 
          |L\!-\!L'|,\ |L\!-\!L'|\!+\!2,\ \ldots,\ 
          (L\!+\!L') \right\}
	  & \;{\Scr H}^{N=2}_{l,l,0}\;, \\[1mm]
  \tilde q = k\!-\!l & \left\{ 
          k\!-\!(L\!+\!L'),\ k\!-\!(L\!+\!L')\!+\!2,\ \ldots,\
          k\!-\!|L\!-\!L'| \right\}
	  & \;{\Scr H}^{N=2}_{l,-l-2,2}\;,
  \end{array}
\end{eqnarray}
where the states of ${\Scr H}^{N=2}_{l,l,0}$ have opposite fermion
number parity as compared with the states of ${\Scr H}^{N=2}_{l,-l-2,2}$. 
This spectrum coincides precisely with the one listed in Tables 1 and 2,
as obtained from the unperturbed Landau-Ginzburg theory; the
label $L$ of the BCFT formulation corresponds to $\Ell$ in the
LG formulation.

%
%
%
%

\vfill\break

\begingroup\raggedright\endgroup

\end{document}